%% file: physa.tex
\documentclass[12pt]{article} 
\usepackage[normal]{caption}
\usepackage[normal]{subfigure}
\usepackage{amsmath}
\usepackage{amssymb}
\usepackage{citesort}
\usepackage{epsfig}
\newcommand{\eq}[1]{Eq.~(\ref{#1})}
\newcommand{\fig}[1]{Fig.~\ref{#1}}
\begin{document}
\title{Ballistic Motion in Quenched Random Environments} 
\author
{Sune Jespersen$^a$\footnote{Corresponding author. Address as
above. Email: sune@ifa.au.dk. Telephone (+45) 89423680, fax (+45)
86120740.}\hspace{.1cm} and
Hans C. Fogedby$^{a,b}$\\ $^a$Institute of Physics and 
Astronomy, University of Aarhus, 8000 {\AA}rhus C, Denmark,\\
$^b$NORDITA, Blegdamsvej 17, 2100 K{\o}benhavn \O, Denmark
}
\date{\today}
\maketitle
\begin{abstract}
We study the motion of a massive particle in a quenched random
environment at zero temperature. The distribution of particle positions
is investigated numerically and special focus is placed on the
mean stopping distance and its fluctuations. We apply
a scaling analysis in order to obtain analytical information about the
distribution function. The model serves as a simple 
example of transport in a random medium.
\end{abstract}
\parbox{1\linewidth}{PACS numbers: 36.20.-r, 47.55.-t, 05.60.-k}
\section{Introduction}
%%%%%%%%%%%%%%%%%%%%%%%%%%%%%%%%%
%%%%%%%%%%%%%%%%%%%%%%%%%%%%%%%%%
%%%%%%%%%%%%%%%%%%%%%%%%%%%%%%%%%
%%%%%%%%%%%%%%%%%%%%%%%%%%%%%%%%%
Dynamics in random media constitute a set of phenomena widely studied 
in modern
statistical physics and soft condensed matter. The problems in questions
extend from the growth of interfaces to the
diffusion of scalars in disordered materials
\cite{sinai,bouchaud1,bouchaud2,bruinsma,koplik,degennes}. In order to 
understand the motion
of manifolds in random environments with focus on the pinning 
mechanism, thermal fluctuations are to a first approximation often
disregarded, see e.g., \cite{fisher1,fisher2,kardar}.
A particular simple case is the motion of a 
massive particle, i.e., a zero dimensional manifold, in a random medium
subject to pinning. 

This model was recently 
analyzed by Stepanow and Schulz \cite{stepanow}. They studied
the behavior of a Newtonian particle 
of mass $m$ in a random environment in $d$-dimensions described by the
equation of motion $m\ddot{\mathbf x}={\mathbf F}({\mathbf
x})$; here ${\mathbf F}$ is 
a quenched random force with a Gaussian
distribution of strength $\sigma$. 
In dimensions $d$ larger than one, the large time results of Stepanow
and Schulz for the first two cumulants are given by 
\begin{equation}
\langle {\mathbf x} \rangle_F\propto 
\frac{v_0^4m^2}{\sigma (d-1)}\hat{{\mathbf v}}_0 ~~~\mbox{and}~~~ 
\langle ({\mathbf x}-\langle{\mathbf x}\rangle_F)^2\rangle_F
\propto t\,\frac{v_0^5m^2d}{\sigma d(d-1)} ~. 
\label{cumulants}
\end{equation}
Here $\hat{{\mathbf v}}_0$ denotes a unit vector along the direction of
the initial velocity ${\mathbf v_0}$. In one
dimension the results in \eq{cumulants} are undefined owing to the
diverging denominators. This behavior reflects the fact that in one
dimension the force $F(x)$ can always be derived from a potential. The
energy is thus 
conserved in this model and the particle will simply
oscillate back and forth between two potential barriers.

In the present paper we extend the results of Stepanow and
Schulz\cite{stepanow} 
by including a friction term in the equation of motion and thus allowing 
for random pinning in the one dimensional case.
The paper is organized as follows. In Section 2 we define the model
and write down the appropriate equations of motion both on the level
of the stochastic equation of motion and the associated Kramers
equation for the time dependent distribution. In Section 3 we
outline the numerical analysis applied to the equation of motion.
In Section 4 we carry out a simple scaling analysis, identifying
the relevant parameters and the scaling functions describing the two
first cumulants of the stationary distribution. Section 5 deals with
the numerical results and we close the paper with a summary and conclusion
in Section 6.  
%%%%%%%%%%%%%%%%%%%%%%%%%%%%%%%%%
%%%%%%%%%%%%%%%%%%%%%%%%%%%%%%%%%
%%%%%%%%%%%%%%%%%%%%%%%%%%%%%%%%%
%%%%%%%%%%%%%%%%%%%%%%%%%%%%%%%%%
\section{The Model}
%%%%%%%%%%%%%%%%%%%%%%%%%%%%%%%%%
%%%%%%%%%%%%%%%%%%%%%%%%%%%%%%%%%
%%%%%%%%%%%%%%%%%%%%%%%%%%%%%%%%%
%%%%%%%%%%%%%%%%%%%%%%%%%%%%%%%%%
In one dimension the motion of a particle of mass $m$ moving in
a viscous medium and subject to a random force
field $F(x)$, depending on the actual position $x$ of the
particle, is governed by the
Langevin equation
\begin{equation}
\label{langevin}
m\frac{d^2x}{dt^2}+\gamma\frac{dx}{dt}=F(x)+\eta(t)~.
\end{equation}
Here $\gamma$ is the coefficient of friction and $\eta(t)$ a thermal
Gaussian white noise with correlations 
$\overline{\eta(t)\eta(t')}=2D\delta(t-t')$. The noise strength $D$
is given by the fluctuation--dissipation theorem, $D=\gamma k_B
T$; $T$ is the temperature of the medium. The distribution of 
$F$ is also chosen to be Gaussian  
with the correlations
\begin{eqnarray}
\langle F(x) \rangle =0 ~~~\mbox{and}~~~  \langle F(x)F(x') \rangle
= \sigma \delta (x-x')~;
\label{moments}
\end{eqnarray}
here $\sigma$ is the force correlation strength.

In the present context we consider the case where the random 
trapping dominates the physics and we shall thus disregard
the thermal fluctuations relative to the random environment
characterized by $F$,  that is  we set 
$D=0$ in \eq{langevin}; this limit corresponds to zero temperature or to
a very massive particle 
$m\rightarrow \infty$. Breaking up 
\eq{langevin} it can then also be written as the set of equations
\begin{eqnarray}
\frac{dv}{dt}&=&-\gamma v+F(x)~, 
\\ 
\frac{dx}{dt}&=&v~,
\label{newton}
\end{eqnarray}
where we for convenience have set $m=1$. Equivalently,
we can also choose to discuss the problem in terms of an equation 
for the distribution function $P(x,v,t)$
itself, namely the Kramers equation in the limit of zero diffusion
coefficient \cite{risken}: 
\begin{equation}
\frac{\partial P(x,v,t)}{\partial t}=
\left[\frac{\partial}{\partial v}
(v\gamma-F(x))-v\frac{\partial}{\partial x} \right]P(x,v,t)~.
\label{kramers}
\end{equation}

As earlier stated, in the limit $\gamma=0$ the model described by
Eqs.~(\ref{newton})-(\ref{kramers}) above reduces to the
one discussed by Stepanow and Schulz in \cite{stepanow}. However, here
we allow for dissipation by having a non-zero $\gamma$, and entirely
different physics will arise.
%%%%%%%%%%%%%%%%%%%%%%%%%%%%%%%%%
%%%%%%%%%%%%%%%%%%%%%%%%%%%%%%%%%
%%%%%%%%%%%%%%%%%%%%%%%%%%%%%%%%%
%%%%%%%%%%%%%%%%%%%%%%%%%%%%%%%%%
\section{Discretization and Numerical Accuracy}
%%%%%%%%%%%%%%%%%%%%%%%%%%%%%%%%%
%%%%%%%%%%%%%%%%%%%%%%%%%%%%%%%%%
%%%%%%%%%%%%%%%%%%%%%%%%%%%%%%%%%
%%%%%%%%%%%%%%%%%%%%%%%%%%%%%%%%%
In order to render the problem amenable to numerical analysis we
modify the force correlations \eq{moments} defined on the continuum
by applying a finite width $a_0$ to the delta function 
(or equivalently, formulating the problem on
a lattice) in such a manner that
\begin{equation}
\langle F(x)F(x') \rangle = \left(\sigma/a_0\right)
\,\delta_{n(x),n(x')} ~.
\label{delta}
\end{equation}
The subscript $n(x)$ of the Kronecker
delta in \eq{delta} is thus a counter of which cell in the lattice the
position $x$ refers to; more precisely,
$n(x)=\left\{ n\in Z \mid x\in[na_0;(n+1)a_0]\right\}$.
In the limit $a_0\to 0$ we recover the full continuum
description in \eq{moments}. 

The equations of motion in \eq{newton} are solved numerically by means
of the fourth order Runge--Kutta method (RK4)
\cite{C}. First, by calling a random generator an actual realization
of the force field is generated corresponding to a specific 
environment.
In the next step we
solve the equations of motion with the specific initial conditions
that the particle initiates its path at $x=0$ at $t=0$ with 
initial velocity $v_0$.
A typical trajectory originating from a simulation is shown in
\fig{trajectory} below. Finally, we
perform an average over realizations of the force field, typically of
the order of 10000 realizations.  

There are three different sources of uncertainty in the simulations:
The finite time step $\tau$, the finite lattice distance $a_0$,
and the finite number of samples of the
random force distribution. Moreover, they are not independent:
Decreasing $a_0$ will require a smaller $\tau$ in order to keep
the error small, since a smaller
lattice constant leads to a more erratic potential in which the
numerical solution is more sensitive to the magnitude of $\tau$.
Choosing a value for $a_0$ we fix $\tau$ such that the stopping distance
of the particle in a specific environment does not change more than a
prescribed fraction (numerical precision) when decreasing $\tau$ by a
factor of say $10$. Then we sample a large number of realizations of the
environment such that the statistical errors are below the desired
precision. The typical relative numerical error in the simulations is about
$1\%$, and the typical relative statistical error is no more than
$5\%$.

%%%%%%%%%%%%%%%%%%%%%%%%%%%%%%%%%
%%%%%%%%%%%%%%%%%%%%%%%%%%%%%%%%%
%%%%%%%%%%%%%%%%%%%%%%%%%%%%%%%%%
%%%%%%%%%%%%%%%%%%%%%%%%%%%%%%%%%
\section{Scaling Analysis}
%%%%%%%%%%%%%%%%%%%%%%%%%%%%%%%%%
%%%%%%%%%%%%%%%%%%%%%%%%%%%%%%%%%
%%%%%%%%%%%%%%%%%%%%%%%%%%%%%%%%%
%%%%%%%%%%%%%%%%%%%%%%%%%%%%%%%%%
Before we embark on a more detailed discussion of the 
numerical results it is
illuminating to perform a simple scaling analysis.
Rescaling space and time in an affine manner according
to $x\rightarrow xb=\tilde{x}$ and $t\rightarrow at=\tilde{t}$,
where $a$ and $b$ are scale parameters to be determined, the 
velocity scales like $v\rightarrow vb/a=\tilde{v}$ and we
obtain by insertion in \eq{newton} the scaled equation of
motion 
$d\tilde{v}/d\tilde{t}=bF(x)F(x)/a^2-(\gamma/a)\tilde{v}$.
The force on the rescaled lattice is thus given by 
$\tilde{F}(\tilde{x})=bF(x)a^2$, is also Gaussian and correlated 
according to 
$\langle\tilde{F}(\tilde{x})\tilde{F}(\tilde{x}')\rangle=
b^3/a^4\sigma\delta(\tilde{x}-\tilde{x}')$. Concluding, we infer the following
scaling of the parameters: 
$\sigma\rightarrow\tilde{\sigma}=\sigma b^3/a^4$, 
$\gamma\rightarrow\tilde{\gamma}=\gamma/a$,
$v_0\rightarrow\tilde{v_0}=(b/a)v_0$,
and
$a_0\rightarrow\tilde{a_0}=ba_0$. Choosing the scale factors $a$ and $b$ 
in such a manner as to
eliminate the dependence on the force strength $\sigma$ and the
friction coefficient $\gamma$, i.e., $a=\gamma$ and 
$b=\gamma^{4/3}/\sigma^{1/3}$, we finally obtain 
the scaling form
\begin{equation}
x(t)=\frac{\sigma^{1/3}}{\gamma^{4/3}}G_1\left(\gamma
t,v_0(\gamma/\sigma)^{1/3},a_0(\gamma^4/\sigma)^{1/3}\right)~,
\end{equation}
where $G_1$ is a random function depending on the actual force
realization. The above scaling analysis can, of course, also
be carried out on the basis of the Kramers equation 
\eq{kramers}.

In the following we limit our discussion 
to the statistical properties of the final
position of the particle. In this limit the dependence on time disappears
and we obtain
\begin{equation}
x\equiv
x(\infty)=
\frac{\sigma^{1/3}}{\gamma^{4/3}}
G\left(v_0\left(\gamma/\sigma\right)^{1/3},a_0(\gamma^4/\sigma)^{1/3}\right)~,
\label{scaling}
\end{equation}
where $G$ is a new scaling function.
\sloppy
The physical meaning of the first argument in the scaling function
$G$ is related to energy
considerations. In one dimension the force $F$ derives from
a potential $U$ according to $U(x)=-\int_0^xF(x')\,dx'$ and, 
correspondingly, the mean square fluctuation of $U$
is given by 
$\langle U^2(x)\rangle=\int_0^x dx'\int_0^x dx''\,\langle
F(x')F(x'')\rangle =\sigma x$.
\fussy
Consequently, the  typical potential energy of the disorder
behaves as $\sqrt{\sigma x}$; this also follows from the fact that
the potential performs a `random walk' in $x$-space. Evaluating this energy
at the trapped position of a  free particle,
i.e., $v_0/\gamma$, we obtain a typical upper limit for the pinning energy.
Comparing this energy to the initial energy of
a free particle, $\sim v_0^2$, we arrive at the following parameter 
characterizing the degree of disorder:
\begin{equation}
\label{parameter}
\mu=\frac{E_{\text{free}}}{E_{\text{disorder}}}\sim
\left(v_0\left(\frac{\gamma}{\sigma}\right)^{1/3}\right)^{3/2}~,
\end{equation}
i.e., a power of the first argument in the scaling function. 
Here small values of the dimensionless parameter $\mu$
corresponds to {\it strong} disorder, whereas large values
of $\mu$ characterizes {\it weak} disorder.
%%%%%%%%%%%%%%%%%%%%%%%%%%%%%%%%%%%%
%%%%%%%%%%%%%%%%%%%%%%%%%%%%%%%%%%%%
%%%%%%%%%%%%%%%%%%%%%%%%%%%%%%%%%%%%
%%%%%%%%%%%%%%%%%%%%%%%%%%%%%%%%%%%%
%%%%%%%%%%%%%%%%%%%%%%%%%%%%%%%%%%%%
\section{Analysis of Numerical Data}
%%%%%%%%%%%%%%%%%%%%%%%%%%%%%%%%%%%%
%%%%%%%%%%%%%%%%%%%%%%%%%%%%%%%%%%%%
%%%%%%%%%%%%%%%%%%%%%%%%%%%%%%%%%%%%
%%%%%%%%%%%%%%%%%%%%%%%%%%%%%%%%%%%%
%%%%%%%%%%%%%%%%%%%%%%%%%%%%%%%%%%%%
In this section we study 
the various moments of the 
variable $x=x(\infty)$ as a function of
the initial velocity $v_0$ and the disorder
strength $\sigma$. Numerical accuracy and the associated computation time 
puts a limit to the range
of values for which we examine the scaling function $G$. We 
have in particular analyzed  the weak disorder regime 
determined by $10\lesssim\mu\lesssim
100$. Stronger disorder requires better routines or much longer
computation time in order to acquire the same accuracy.
%%%%%%%%%%%%%%%%%%%%%%%%%%%%%%%%%%%%
%%%%%%%%%%%%%%%%%%%%%%%%%%%%%%%%%%%%
%%%%%%%%%%%%%%%%%%%%%%%%%%%%%%%%%%%%
%%%%%%%%%%%%%%%%%%%%%%%%%%%%%%%%%%%%
%%%%%%%%%%%%%%%%%%%%%%%%%%%%%%%%%%%%
\subsection{The trajectory}
%%%%%%%%%%%%%%%%%%%%%%%%%%%%%%%%%%%%
%%%%%%%%%%%%%%%%%%%%%%%%%%%%%%%%%%%%
%%%%%%%%%%%%%%%%%%%%%%%%%%%%%%%%%%%%
%%%%%%%%%%%%%%%%%%%%%%%%%%%%%%%%%%%%
%%%%%%%%%%%%%%%%%%%%%%%%%%%%%%%%%%%%
It is instructive first to consider the outcome of a typical
simulation. In Fig. 1
%\ref{trajectory} 
we have thus shown
a simulated trajectory with a superposed trajectory of 
the free flight, corresponding to the solution
of \eq{newton} with
$F(x)=0$, i.e.,
\begin{equation}
x_{\text{free}}(t)=\frac{v_0}{\gamma}(1-\exp({-\gamma t}))~,
\label{free}
\end{equation}
with $x(0)=0$ and $v(0)=v_0$.

The initial motion follows that of a free particle, however, as the kinetic
energy degrades due to friction the 
particle will soon experience a (random) force large enough to 
reverse the
motion. In order to illustrate this effect we depict Fig.~\ref{energy}.a
the potential energy
$U(x)$ (solid curve) and the kinetic energy of free flight 
(the dashed curve)
a function of position. The point of intersection of the two curves 
is to a first
approximation  the point of first reflection of the
particle.  
The actual motion of the particle is, of course,
sensitive to the disorder before the point of reflection,
perturbing the dashed curve on the figure. After the first reflection
the velocity is reversed and the process reiterates. Owing to the
gradual loss of kinetic energy the particle travels shorter and
shorter distances and finally comes to  rest in a local minimum of the
potential. Figure~\ref{energy}.b is a magnification of the region around
the minimum that finally traps the particle.

In Fig.~\ref{dist}.a we show the distribution of final
positions and note the general trend that the particle travels shorter
than for the corresponding free motion. This is illustrated by the
vertical line on the figure, indicating  the final position for a
free particle with the same initial velocity.
Despite a superficial resemblance the distribution depicted in
Fig.~\ref{dist}.a is not Gaussian. A transparent way to exhibit this
deviation from Gaussian behavior is to plot the logarithm of the distribution;
a Gaussian distribution then yields a parabolic shape. This plot is shown
in Fig. \ref{dist}.b, where  we have also included a fit to a
parabola. It follows from the fit that the actual distribution 
possesses long tails.

%(COMMENT:
%Do we have law for tails of distribution, rare events??)
%%%%%%%%%%%%%%%%%%%%%%%%%%%%%%%%%%%%%%%%%%%
%%%%%%%%%%%%%%%%%%%%%%%%%%%%%%%%%%%%%%%%%%%
%%%%%%%%%%%%%%%%%%%%%%%%%%%%%%%%%%%%%%%%%%%
%%%%%%%%%%%%%%%%%%%%%%%%%%%%%%%%%%%%%%%%%%%
%%%%%%%%%%%%%%%%%%%%%%%%%%%%%%%%%%%%%%%%%%%
%%%%%%%%%%%%%%%%%%%%%%%%%%%%%%%%%%%%%%%%%%%
\subsection{Mean distance traveled}
%%%%%%%%%%%%%%%%%%%%%%%%%%%%%%%%%%%%%%%%%%%
%%%%%%%%%%%%%%%%%%%%%%%%%%%%%%%%%%%%%%%%%%%
%%%%%%%%%%%%%%%%%%%%%%%%%%%%%%%%%%%%%%%%%%%
%%%%%%%%%%%%%%%%%%%%%%%%%%%%%%%%%%%%%%%%%%%
%%%%%%%%%%%%%%%%%%%%%%%%%%%%%%%%%%%%%%%%%%%
%%%%%%%%%%%%%%%%%%%%%%%%%%%%%%%%%%%%%%%%%%%
A central quantity is the total mean distance traveled by
a particle, $\langle x\rangle_F$, as a function of the initial
velocity $v_0$ and the force correlation strength
$\sigma$. The notation $\langle\cdot\rangle_F$ here denotes
an average over the Gaussian force distributions.
In \fig{meanvss} we have depicted $\langle x\rangle_F$
as a function of $\sigma$ for fixed damping $\gamma=5.0$ and 
initial velocity $v_0=10.0$.
Data from two different seeds, i.e., two series of realizations of the
random force field, are shown in order to indicate 
the statistical uncertainty.

%(COMMENT: Is it possible to plot error bars)

The data are reasonably independent of the small
distance cutoff $a_0$, and thus applying the scaling form
\eq{scaling} we infer
\begin{equation}
\langle x \rangle_F=\frac{\sigma^{1/3}}{\gamma^{4/3}}\left\langle
G(v_0(\gamma/\sigma)^{1/3})\right\rangle_F~.
\label{mean}
\end{equation}
A good fit to the data in \fig{meanvss} is provided by the
stretched exponential
$\langle x\rangle_F\sim a\exp\left({-b\sigma^c}\right)$.
Averaging over a set of data with different cutoffs $a_0$ 
in the range from $0.01$ to $0.001$ we obtain for the 
parameters $a=2.000\pm 0.001$, $b=0.058\pm0.001$, and
$c=0.34\pm 0.01$. The error bars are extracted from an
appropriate sample of data. The data are thus consistent with
the scaling form
\begin{equation}
\langle x\rangle_F\sim\frac{v_0}{\gamma}
\exp\left({-\frac{(\sigma/\gamma)^{1/3}}{v_0}}\right)~,
\label{G1}
\end{equation}
which has the correct limiting form $x\sim v_0/\gamma$ in the
absence of disorder or large initial velocity $v_0$. Also, in the
limit of vanishing friction $\gamma\rightarrow 0$ the first
moment $\langle x\rangle_F$ vanishes. Consequently, there is
a largest mean distance traveled for an appropriately 
selected value of the friction. Since we are here considering the weak
disorder regime one could argue that the first two terms in the
expansion of the stretched exponential are sufficient.
We shall however stick to the expression in
\eq{G1} since it behaves correctly in the limit $\gamma\to 0$. 

As an alternative check of the scaling form \eq{G1} we next
vary the initial velocity $v_0$. The data are shown in 
\fig{meanvsv} for the values
$\gamma=5.0$ and $\sigma=1.0$. It follows that \eq{G1} does describe
the data well. Also shown is a linear fit and a fit to a
stretched exponential. We note that all curves appear to coincide
on the figure, thus not allowing us to distinguish the best fit.

%(MORE COMMENTS)
%%%%%%%%%%%%%%%%%%%%%%%%%%%%%%%%%%%%%%%%%%%
%%%%%%%%%%%%%%%%%%%%%%%%%%%%%%%%%%%%%%%%%%%
%%%%%%%%%%%%%%%%%%%%%%%%%%%%%%%%%%%%%%%%%%%
%%%%%%%%%%%%%%%%%%%%%%%%%%%%%%%%%%%%%%%%%%%
%%%%%%%%%%%%%%%%%%%%%%%%%%%%%%%%%%%%%%%%%%%
%%%%%%%%%%%%%%%%%%%%%%%%%%%%%%%%%%%%%%%%%%%
\subsection{Fluctuations in the stopping distance}
%%%%%%%%%%%%%%%%%%%%%%%%%%%%%%%%%%%%%%%%%%%
%%%%%%%%%%%%%%%%%%%%%%%%%%%%%%%%%%%%%%%%%%%
%%%%%%%%%%%%%%%%%%%%%%%%%%%%%%%%%%%%%%%%%%%
%%%%%%%%%%%%%%%%%%%%%%%%%%%%%%%%%%%%%%%%%%%
%%%%%%%%%%%%%%%%%%%%%%%%%%%%%%%%%%%%%%%%%%%
%%%%%%%%%%%%%%%%%%%%%%%%%%%%%%%%%%%%%%%%%%%
The last issue we address is the root mean square
fluctuations of the stopping distance
$\Delta x = \sqrt{\langle (x-\langle x\rangle)^2\rangle}$,
depicted in \fig{dist}.
Here the situation is a little complicated since we
find a dependence on the small distance cutoff $a_0$.
From the general scaling analysis yielding \eq{scaling}
we infer a scaling form for $\Delta x$,
\begin{equation}
\Delta x=\sqrt{\langle (x-\langle
x\rangle)^2\rangle}=\frac{\sigma^{1/3}}{\gamma^{4/3}}
G_2\left(v_0(\gamma/\sigma)^{1/3},a_0(\gamma^4/\sigma)^{1/3}\right)~,
\label{G2}
\end{equation}
where $G_2$ is a new unknown scaling function.

In \fig{sdevvss} we have for fixed damping $\gamma=5.0$ and
initial velocity $v_0=10.0$ plotted the standard deviation
of $x$ versus the disorder strength $\sigma$ for two
different cutoffs $a_0$ on a log-log scale. The linear fit
suggests the simple power law behavior 
\begin{equation}
\Delta x=u_1\sigma^{z_1}.
\label{powerlaw}
\end{equation}
In order to determine the dependence on the lattice 
constant $a_0$ we have in Figs.~\ref{expvsa} and \ref{ampvsa}
plotted the exponent $z_1$ and the amplitude $u_1$ as
a function of $a_0$ in order
to study the limiting function
$G_2(v_0(\gamma/\sigma)^{1/3})=
\lim_{a_0\rightarrow 0}
G_2(v_0(\gamma/\sigma)^{1/3},a_0(\gamma^4/\sigma)^{1/3})$
for decreasing $a_0$. With a linear fit we find the following values
for the amplitude and the exponent:
\begin{eqnarray}
u_1=0.143\pm 0.001,\hspace{.3cm}z_1=0.326\pm 0.002,
\end{eqnarray}
indicating that the best fit is obtained using a constant
scaling function $G_2 = \text{const.}$. We are thus led to the
form
\begin{equation}
\Delta x=\text{const.}\frac{\sigma^{1/3}}{\gamma^{4/3}}~,
\label{ansatz}
\end{equation}
independent of initial velocity. Substantiating our result we
have finally in \fig{sdevvsv} plotted $\Delta x$ versus $v_0$. The
independence of $\Delta x$ on $v_0$ is quite convincing. 
%%%%%%%%%%%%%%%%%%%%%%%%%%%%%%%%%%%%%%%%%%%%%%
%%%%%%%%%%%%%%%%%%%%%%%%%%%%%%%%%%%%%%%%%%%%%%
%%%%%%%%%%%%%%%%%%%%%%%%%%%%%%%%%%%%%%%%%%%%%%
%%%%%%%%%%%%%%%%%%%%%%%%%%%%%%%%%%%%%%%%%%%%%%
%%%%%%%%%%%%%%%%%%%%%%%%%%%%%%%%%%%%%%%%%%%%%%
%%%%%%%%%%%%%%%%%%%%%%%%%%%%%%%%%%%%%%%%%%%%%%
\section{Summary and Conclusions}
%%%%%%%%%%%%%%%%%%%%%%%%%%%%%%%%%%%%%%%%%%%%%%
%%%%%%%%%%%%%%%%%%%%%%%%%%%%%%%%%%%%%%%%%%%%%%
%%%%%%%%%%%%%%%%%%%%%%%%%%%%%%%%%%%%%%%%%%%%%%
%%%%%%%%%%%%%%%%%%%%%%%%%%%%%%%%%%%%%%%%%%%%%%
%%%%%%%%%%%%%%%%%%%%%%%%%%%%%%%%%%%%%%%%%%%%%%
%%%%%%%%%%%%%%%%%%%%%%%%%%%%%%%%%%%%%%%%%%%%%%
In this paper we have studied a 
simple model for damped ballistic transport in
a random environment: A massive particle moving in a random
medium also subject to a viscous force. From
energy considerations alone it is clear that the particle will
eventually become pinned at a local minimum of the potential and we
have thus restricted our attention to the properties of the stationary
distribution. We performed an elementary scaling analysis of the
equation of motion and exploited the results in the ensuing numerical
analysis. Here we restricted our attention to the behavior of the
first two cumulants, i.e. the mean stopping distance and the 
mean square fluctuations of the stopping distance
as functions of disorder strength and initial velocity.
We found that the data in the weak disorder regime were
well described by the expressions:
\begin{equation}
\langle
x\rangle_F\sim\frac{v_0}{\gamma}\exp\left({-\frac{(\sigma/\gamma)^{1/3}}{v_0}}\right),
\hspace{.5cm}  \left\langle(x-\langle
x\rangle)^2\right\rangle^{1/2}\sim
\text{const.}\frac{\sigma^{1/3}}{\gamma^{4/3}}\;.
\end{equation}
We, moreover, exhibited the non-Gaussian nature of the stationary
distribution of stopping positions of the particle.

There are many more
quantities of interest that one could extract numerically. First of
all it would be interesting to explore also the dynamic properties, that
is the behavior of e.g. $x(t)$ and $v(t)$. Moreover, a more thorough
analytical examination would also be desirable. Finally, an
extension of the model to the case of a random force with a
non-vanishing mean value would introduce some new facets into the
problem.

\newpage
\noindent
{\large\bf Figures}

\begin{figure}[h]
\unitlength=1cm
\begin{center}
\begin{picture}(6,5.8)
\put(-2.2,6.6)
{ 
\includegraphics{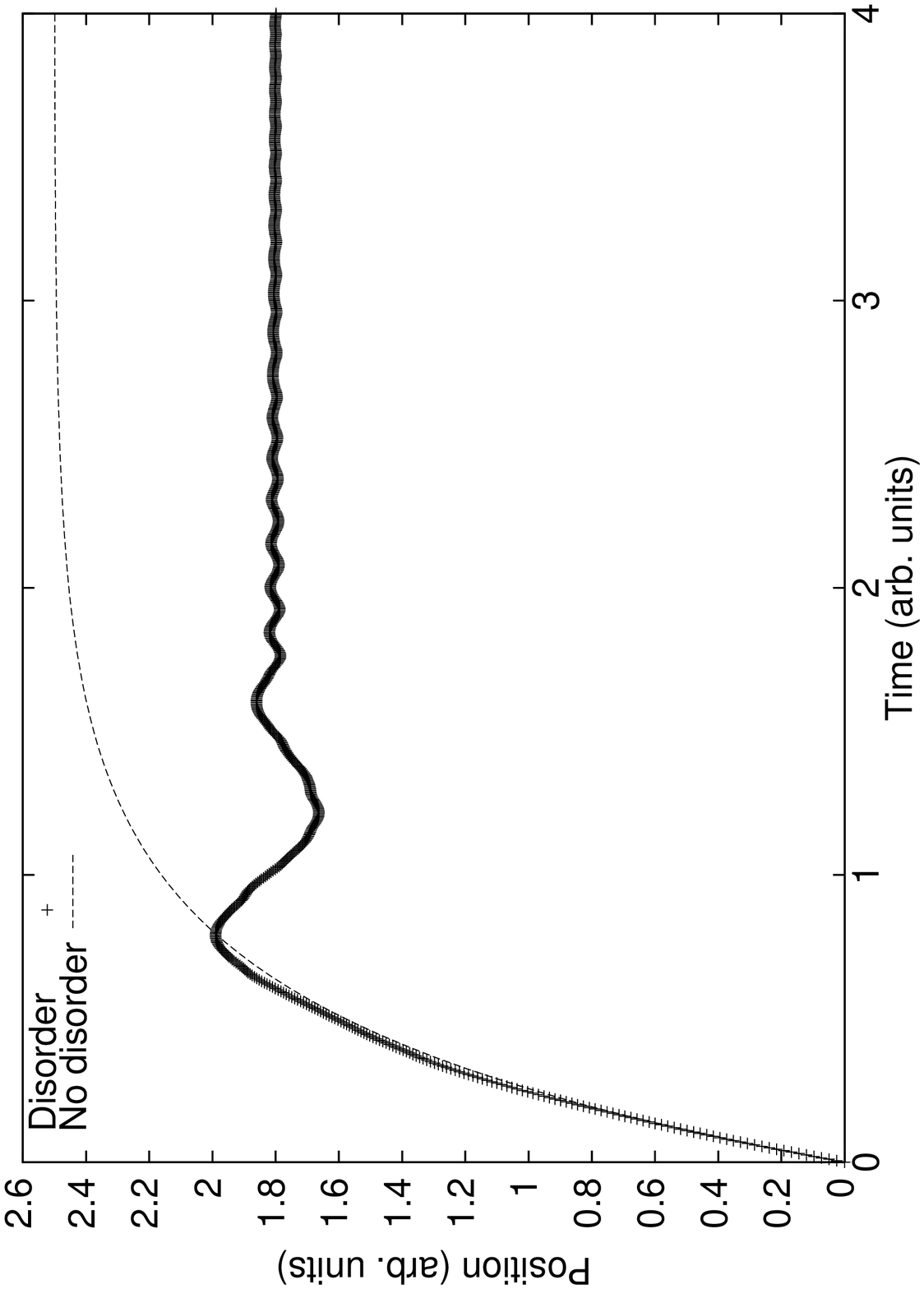}
}
\end{picture}
\end{center}
\caption[]{\label{trajectory}
We show a typical trajectory of ballistic motion in one 
dimension in a random force field (the full line). For comparison 
we also depict
the trajectory of the motion in the absence of disorder 
(the dashed line). Notice how the trajectories are very similar at
short times, whereas for larger times they differ substantially.}
\end{figure} 

\begin{figure}[ht]
\begin{center}
\begin{tabular}{lr}
\subfigure[]{\epsfig{angle=0,figure=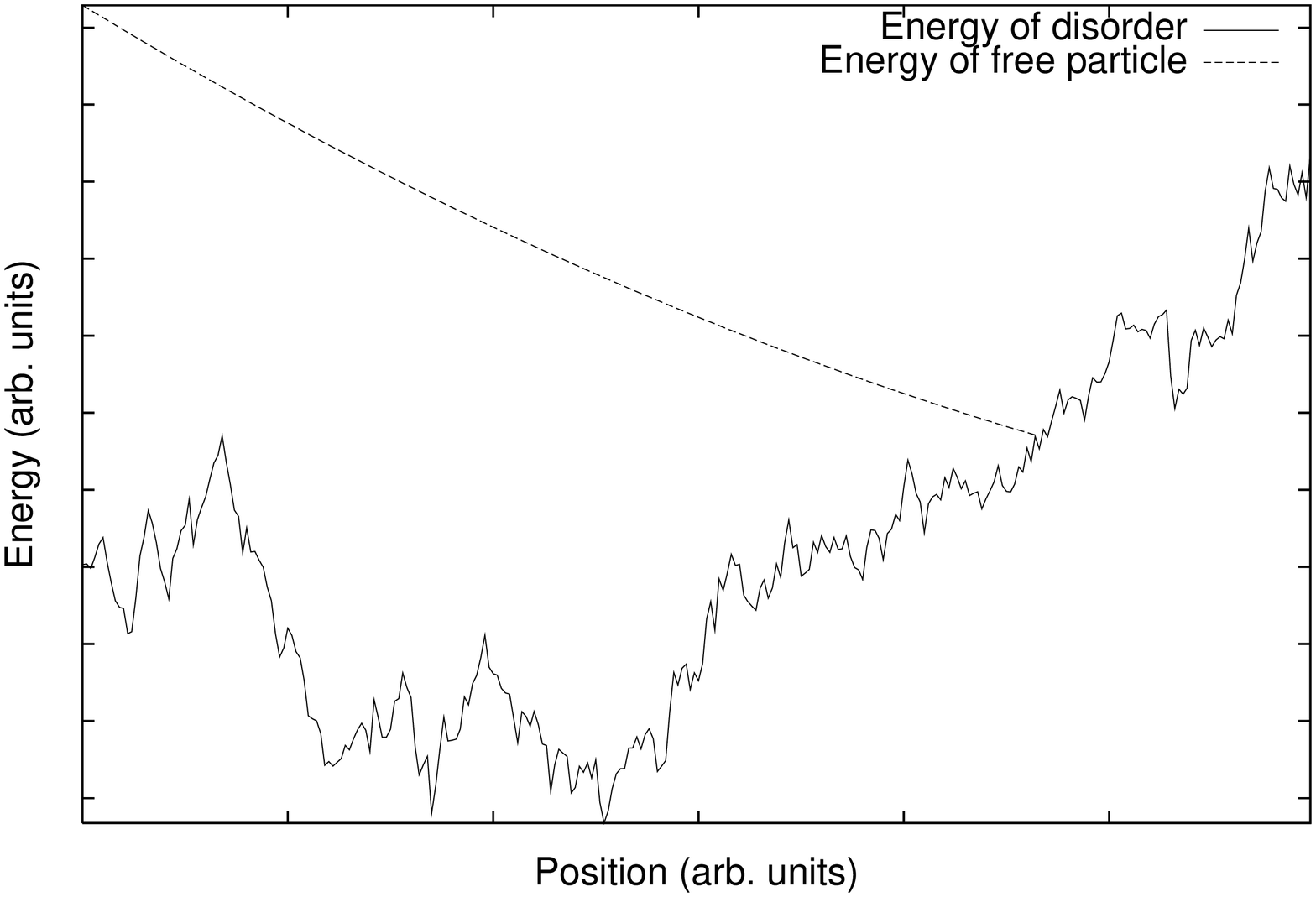,width=.4\textwidth}}&
\subfigure[]{\epsfig{angle=0,figure=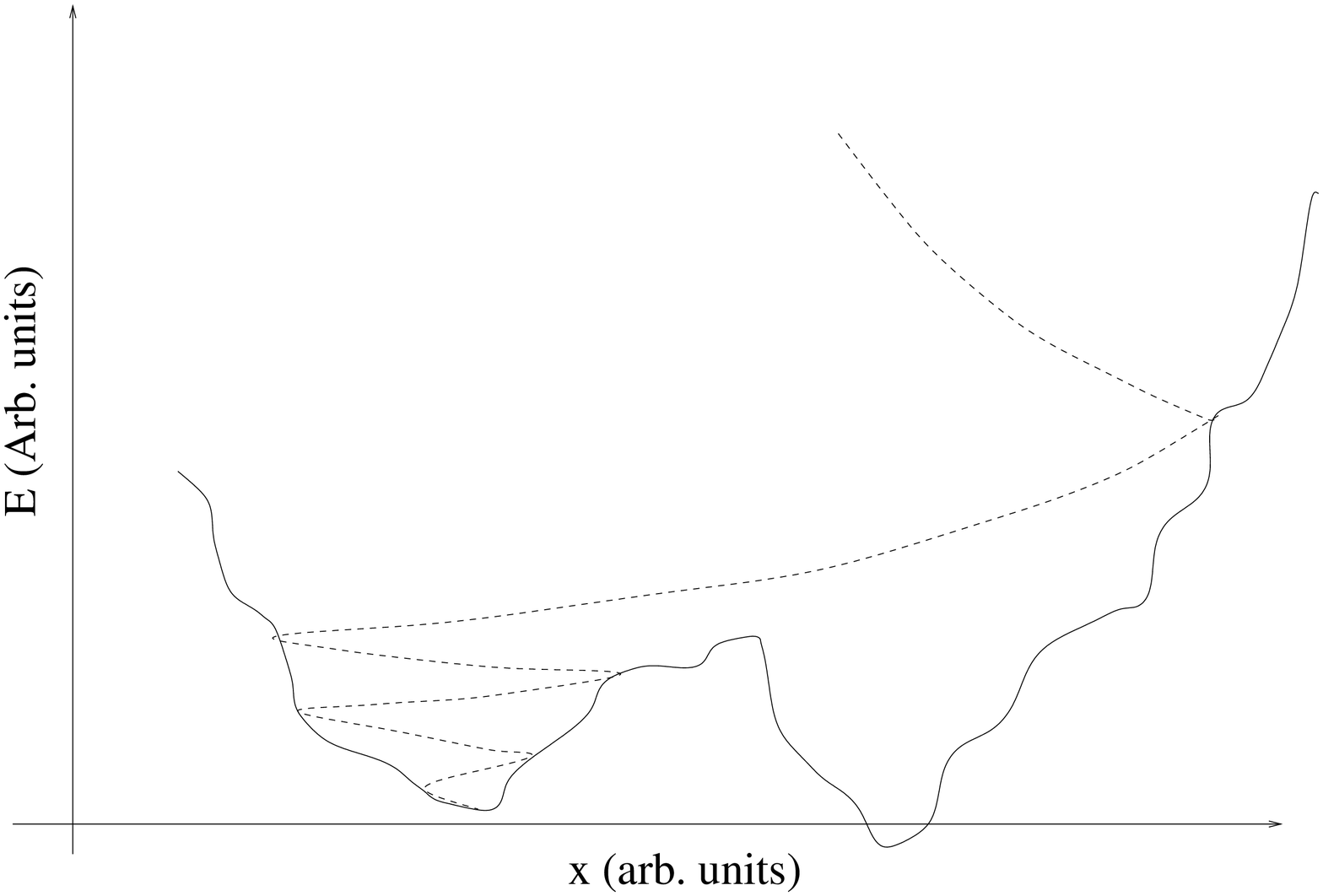,width=.4\textwidth}}
\end{tabular}
\end{center}
\caption{\label{energy}
(a): Comparison of the typical energy of the disorder, which is
a random walk in $x$-space, with the energy of the free particle, as a
function of position. (b): The particle reflects a
number of times on the potential and gradually looses its energy,
until it is trapped in a local minimum of the potential.} 
\end{figure}  

\begin{figure}[ht]
\unitlength=1cm
\begin{center}
\begin{tabular}{lr}
\subfigure[]{\epsfig{figure=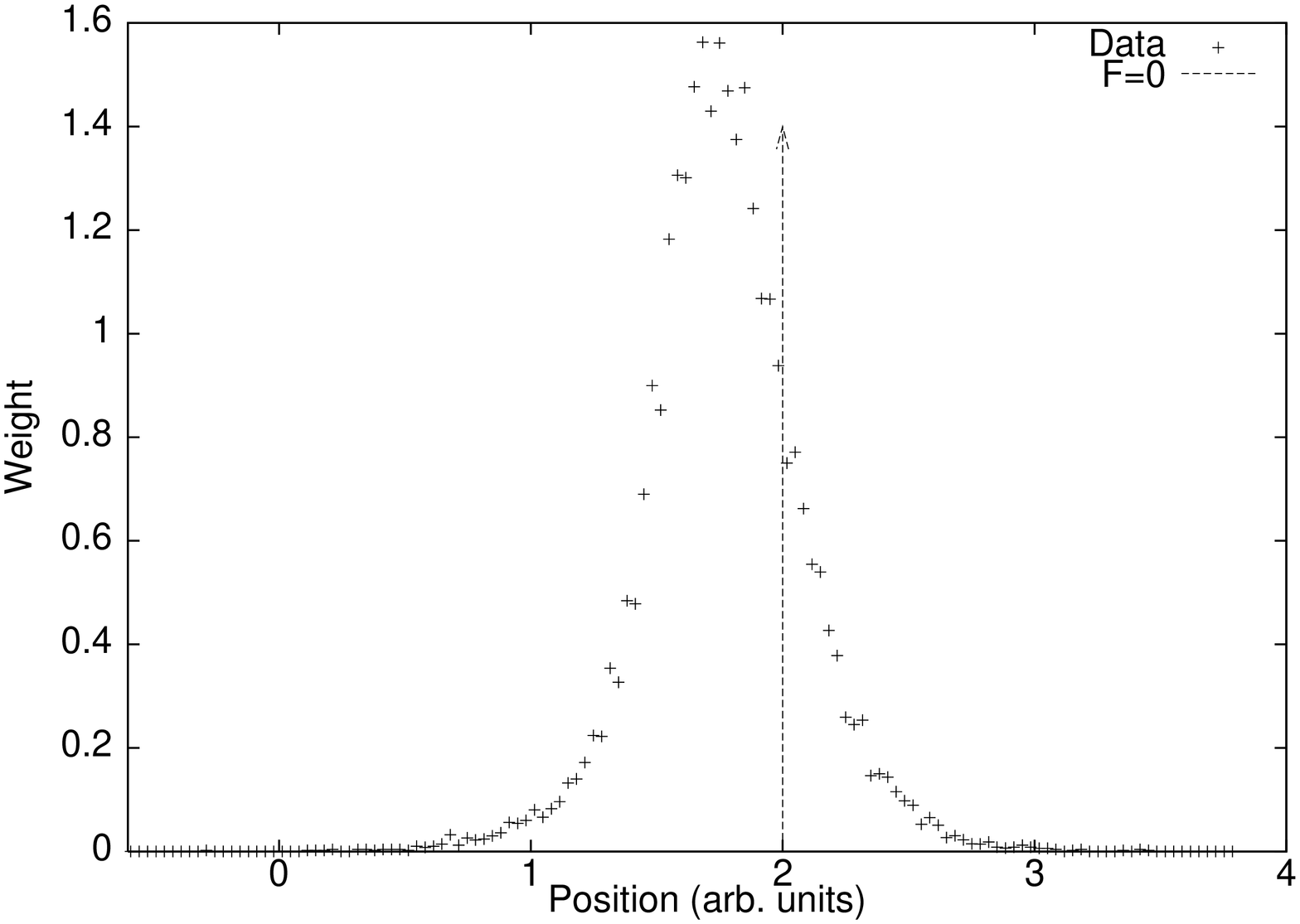,angle=270,width=.4\textwidth}}&
\subfigure[]{\epsfig{figure=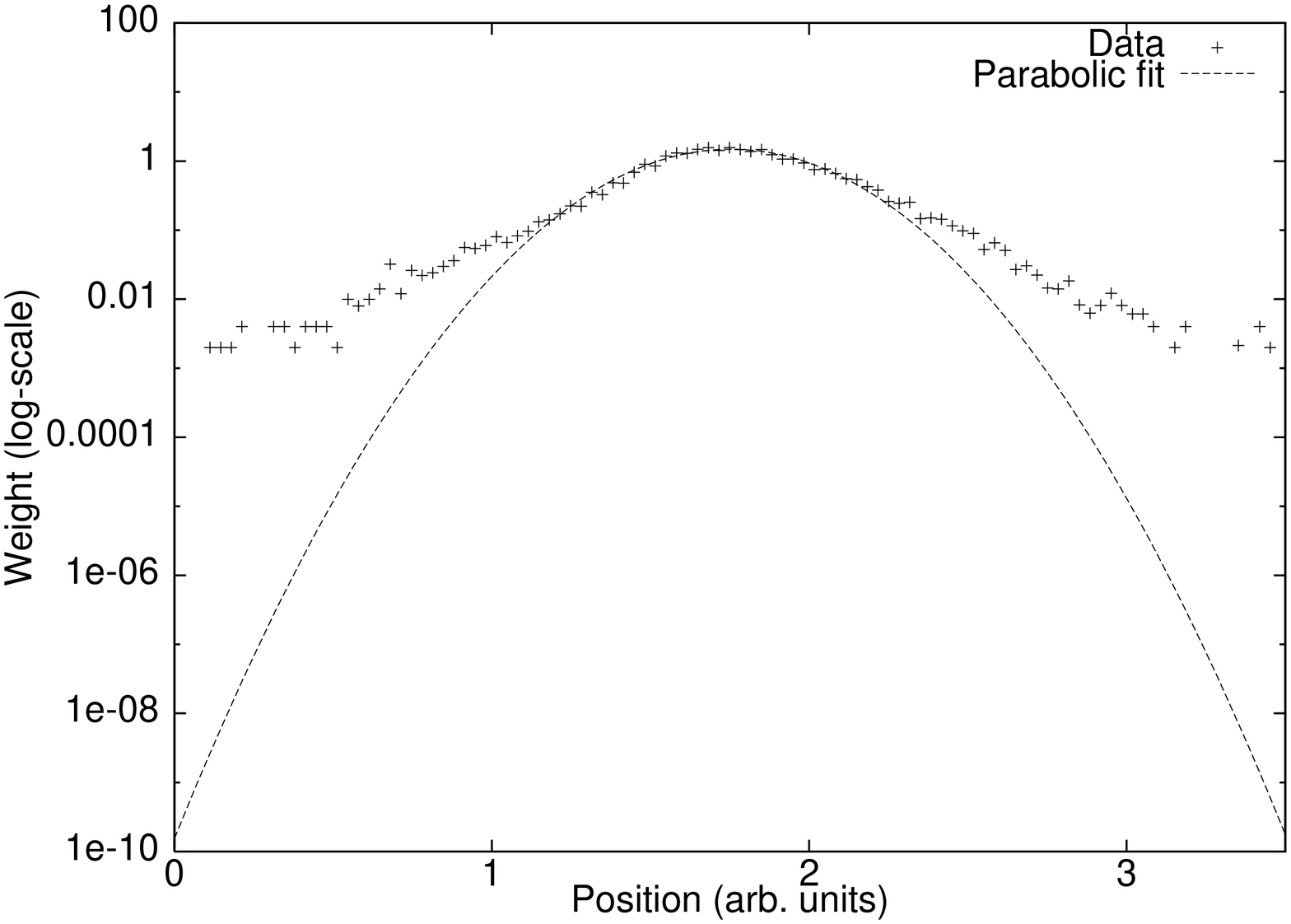,angle=270,width=.4\textwidth}}
\end{tabular}
\end{center}
\caption{\label{dist}
 In (a), we plot the stationary distribution of the particles
rest position.  The vertical line at $x=2$ marks the final position of
a free particle having the same initial velocity. In (b) we have shown
the same data on a semi log scale. A fit to a parabola is also
shown, and the deviation from Gaussian behavior is seen to be large
especially away from the center.}
\end{figure}  

\begin{figure}[ht]
\begin{center}
\input{fig4}
\end{center}
\caption{\label{meanvss}
The mean stopping distance as a function of disorder strength
$\sigma$. The data (from both seeds) have been fitted to a stretched
exponential (solid line).}
\end{figure}
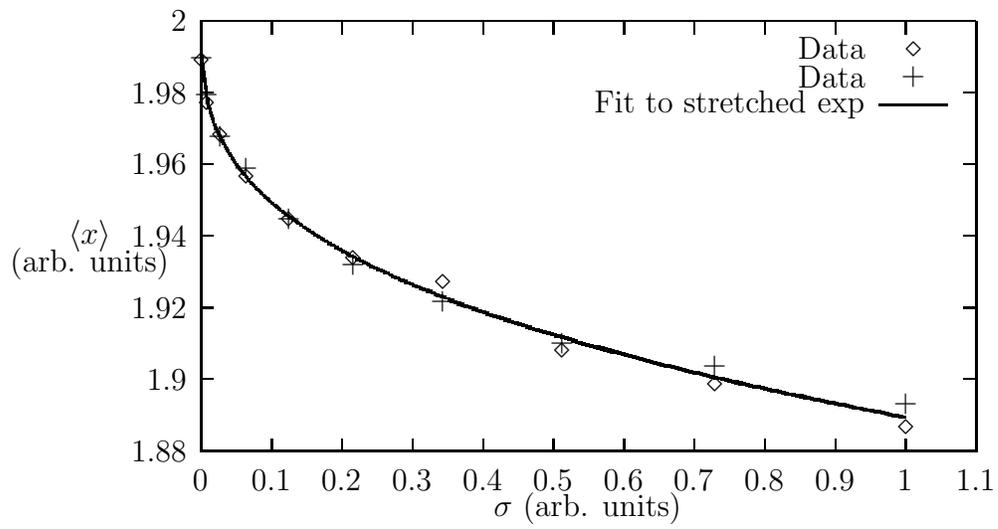  

\begin{figure}[ht]
\begin{center}
\input{fig5}
\end{center}
\caption{\label{meanvsv}
This is a plot of $\langle x\rangle$ versus $v_0$. We have
shown both a linear fit, a fit to a stretched exponential, and also
the prediction \eq{G1}. They are all indistinguishable in this plot.}
\end{figure}
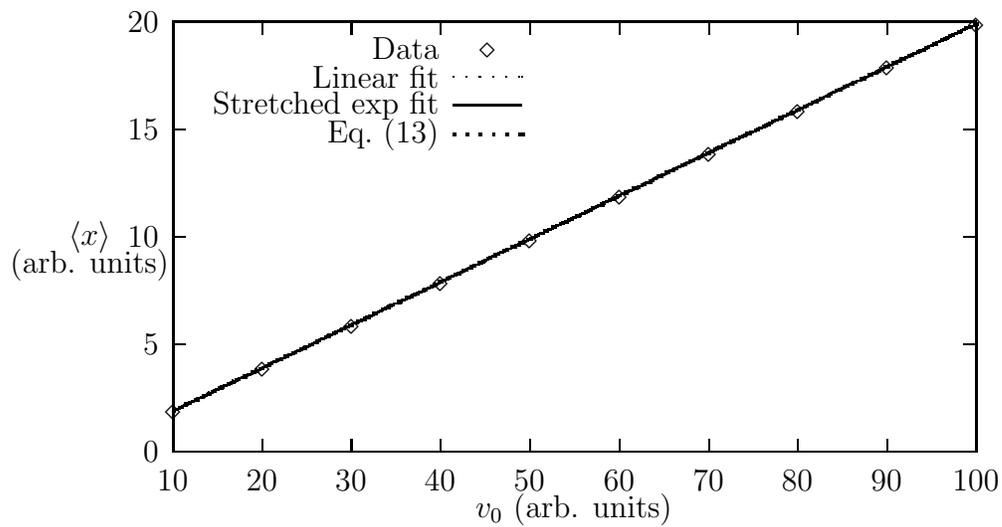

\begin{figure}[ht]
\begin{center}
\input{fig6}
\end{center}
\caption{\label{sdevvss}
A graph of standard deviation $\Delta x$ of stopping distribution as a
function of strength of disorder. The two
different sets of data are for two different lattice constants. The
straight lines indicate power-law behavior, albeit with different
exponents and amplitudes. }
\end{figure}
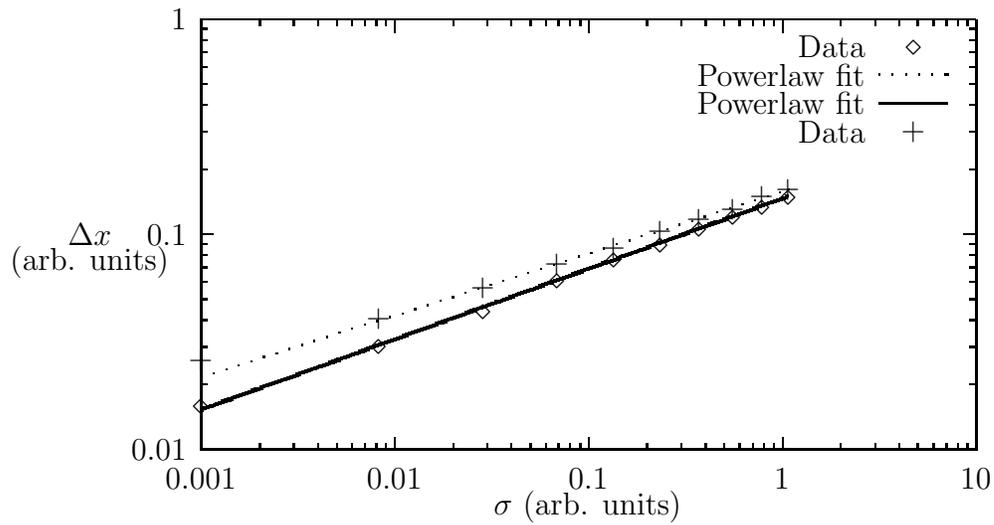

\begin{figure}[ht]
\begin{center}
\input{fig7}
\end{center}
\caption{\label{expvsa}
Exponent $z_1$ (see \eq{powerlaw} of the power
law fits as a function of the lattice constant, and a fit to a
straight line. We are interested in the limit $a_0\to 0$.}
\end{figure}
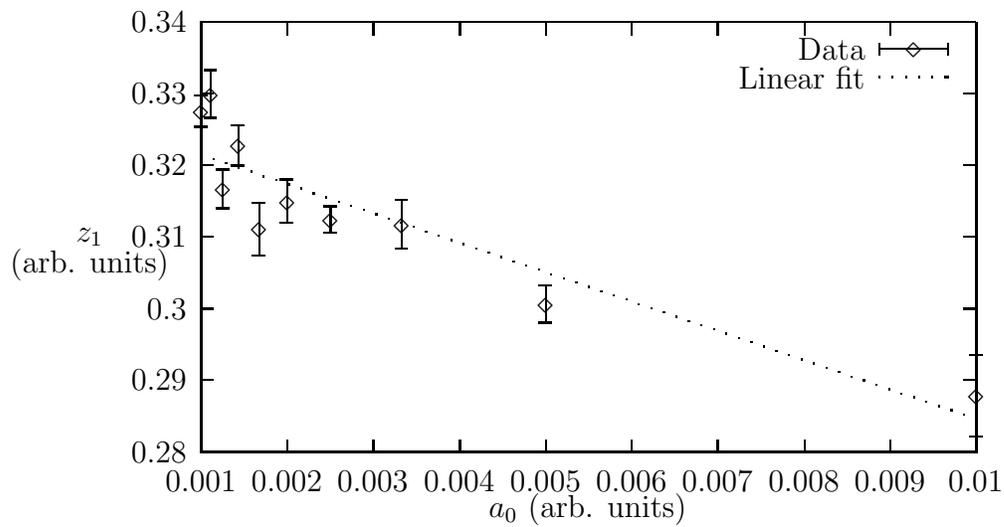

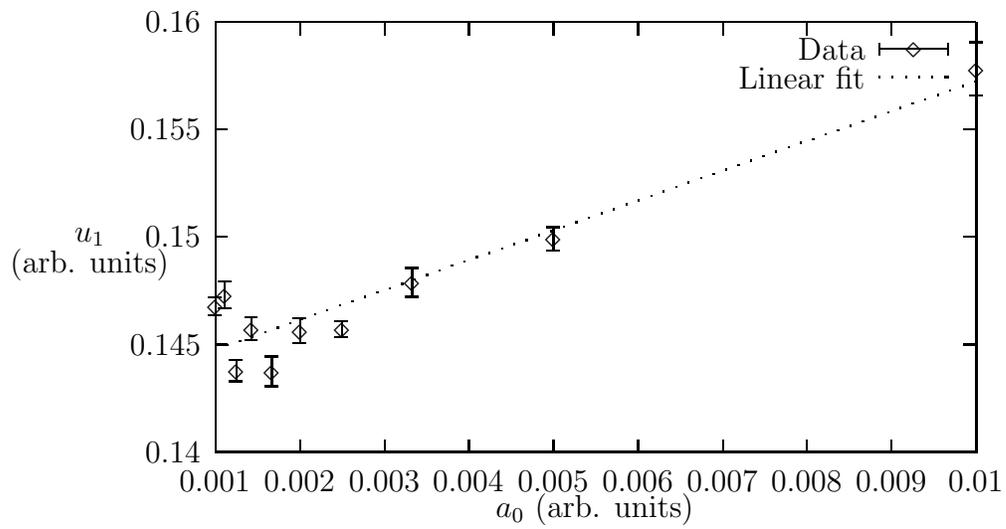
\begin{figure}[ht]
\begin{center}
\input{fig8}
\end{center}
\caption{\label{ampvsa}
Amplitude $u_1$ in \eq{powerlaw} of the power law fits as
a function of the lattice constant, and a fit to a straight line.}
\end{figure}

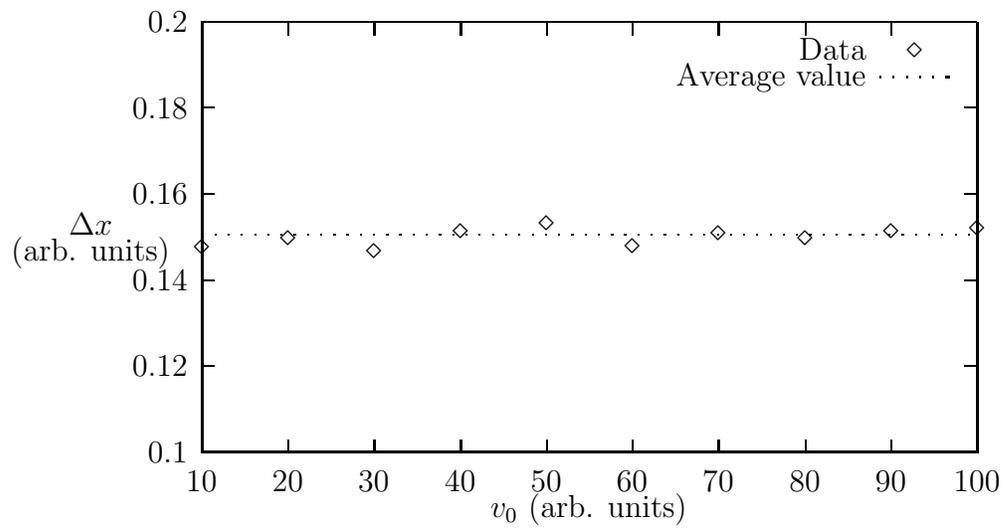
\begin{figure}[ht]
\begin{center}
\input{fig9}
\end{center}
\caption{\label{sdevvsv}
The dependence of the width of stopping distances on the
initial speed $v_0$ is seen to be well approximated by a constant
expression. From this figure we have $\Delta x=0.154$.}
\end{figure}

\end{document}

%% file: fig4.tex
% GNUPLOT: LaTeX picture
\setlength{\unitlength}{0.240900pt}
\ifx\plotpoint\undefined\newsavebox{\plotpoint}\fi
\sbox{\plotpoint}{\rule[-0.200pt]{0.400pt}{0.400pt}}%
\begin{picture}(1500,900)(0,0)
\font\gnuplot=cmr12 at 12pt
\gnuplot
\sbox{\plotpoint}{\rule[-0.200pt]{0.400pt}{0.400pt}}%
\put(219.0,134.0){\rule[-0.200pt]{4.818pt}{0.400pt}}
\put(197,134){\makebox(0,0)[r]{1.88}}
\put(1416.0,134.0){\rule[-0.200pt]{4.818pt}{0.400pt}}
\put(219.0,247.0){\rule[-0.200pt]{4.818pt}{0.400pt}}
\put(197,247){\makebox(0,0)[r]{1.9}}
\put(1416.0,247.0){\rule[-0.200pt]{4.818pt}{0.400pt}}
\put(219.0,359.0){\rule[-0.200pt]{4.818pt}{0.400pt}}
\put(197,359){\makebox(0,0)[r]{1.92}}
\put(1416.0,359.0){\rule[-0.200pt]{4.818pt}{0.400pt}}
\put(219.0,472.0){\rule[-0.200pt]{4.818pt}{0.400pt}}
\put(197,472){\makebox(0,0)[r]{1.94}}
\put(1416.0,472.0){\rule[-0.200pt]{4.818pt}{0.400pt}}
\put(219.0,585.0){\rule[-0.200pt]{4.818pt}{0.400pt}}
\put(197,585){\makebox(0,0)[r]{1.96}}
\put(1416.0,585.0){\rule[-0.200pt]{4.818pt}{0.400pt}}
\put(219.0,697.0){\rule[-0.200pt]{4.818pt}{0.400pt}}
\put(197,697){\makebox(0,0)[r]{1.98}}
\put(1416.0,697.0){\rule[-0.200pt]{4.818pt}{0.400pt}}
\put(219.0,810.0){\rule[-0.200pt]{4.818pt}{0.400pt}}
\put(197,810){\makebox(0,0)[r]{2}}
\put(1416.0,810.0){\rule[-0.200pt]{4.818pt}{0.400pt}}
\put(219.0,134.0){\rule[-0.200pt]{0.400pt}{4.818pt}}
\put(219,89){\makebox(0,0){0}}
\put(219.0,790.0){\rule[-0.200pt]{0.400pt}{4.818pt}}
\put(330.0,134.0){\rule[-0.200pt]{0.400pt}{4.818pt}}
\put(330,89){\makebox(0,0){0.1}}
\put(330.0,790.0){\rule[-0.200pt]{0.400pt}{4.818pt}}
\put(440.0,134.0){\rule[-0.200pt]{0.400pt}{4.818pt}}
\put(440,89){\makebox(0,0){0.2}}
\put(440.0,790.0){\rule[-0.200pt]{0.400pt}{4.818pt}}
\put(551.0,134.0){\rule[-0.200pt]{0.400pt}{4.818pt}}
\put(551,89){\makebox(0,0){0.3}}
\put(551.0,790.0){\rule[-0.200pt]{0.400pt}{4.818pt}}
\put(662.0,134.0){\rule[-0.200pt]{0.400pt}{4.818pt}}
\put(662,89){\makebox(0,0){0.4}}
\put(662.0,790.0){\rule[-0.200pt]{0.400pt}{4.818pt}}
\put(772.0,134.0){\rule[-0.200pt]{0.400pt}{4.818pt}}
\put(772,89){\makebox(0,0){0.5}}
\put(772.0,790.0){\rule[-0.200pt]{0.400pt}{4.818pt}}
\put(883.0,134.0){\rule[-0.200pt]{0.400pt}{4.818pt}}
\put(883,89){\makebox(0,0){0.6}}
\put(883.0,790.0){\rule[-0.200pt]{0.400pt}{4.818pt}}
\put(993.0,134.0){\rule[-0.200pt]{0.400pt}{4.818pt}}
\put(993,89){\makebox(0,0){0.7}}
\put(993.0,790.0){\rule[-0.200pt]{0.400pt}{4.818pt}}
\put(1104.0,134.0){\rule[-0.200pt]{0.400pt}{4.818pt}}
\put(1104,89){\makebox(0,0){0.8}}
\put(1104.0,790.0){\rule[-0.200pt]{0.400pt}{4.818pt}}
\put(1215.0,134.0){\rule[-0.200pt]{0.400pt}{4.818pt}}
\put(1215,89){\makebox(0,0){0.9}}
\put(1215.0,790.0){\rule[-0.200pt]{0.400pt}{4.818pt}}
\put(1325.0,134.0){\rule[-0.200pt]{0.400pt}{4.818pt}}
\put(1325,89){\makebox(0,0){1}}
\put(1325.0,790.0){\rule[-0.200pt]{0.400pt}{4.818pt}}
\put(1436.0,134.0){\rule[-0.200pt]{0.400pt}{4.818pt}}
\put(1436,89){\makebox(0,0){1.1}}
\put(1436.0,790.0){\rule[-0.200pt]{0.400pt}{4.818pt}}
\put(219.0,134.0){\rule[-0.200pt]{293.175pt}{0.400pt}}
\put(1436.0,134.0){\rule[-0.200pt]{0.400pt}{162.848pt}}
\put(219.0,810.0){\rule[-0.200pt]{293.175pt}{0.400pt}}
\put(45,472){\makebox(0,0){$\langle x\rangle$}}
\put(45,428){\makebox(0,0){(arb. units)}}
\put(827,44){\makebox(0,0){$\sigma$ (arb. units)}}
%\put(827,855){\makebox(0,0){Mean stopping distance vs. disorder strength}}
\put(219.0,134.0){\rule[-0.200pt]{0.400pt}{162.848pt}}
\put(1262,768){\makebox(0,0)[r]{Data}}
\put(1338,768){\raisebox{-.8pt}{\makebox(0,0){$\diamond$}}}
\put(220,751){\raisebox{-.8pt}{\makebox(0,0){$\diamond$}}}
\put(228,684){\raisebox{-.8pt}{\makebox(0,0){$\diamond$}}}
\put(249,633){\raisebox{-.8pt}{\makebox(0,0){$\diamond$}}}
\put(290,568){\raisebox{-.8pt}{\makebox(0,0){$\diamond$}}}
\put(357,501){\raisebox{-.8pt}{\makebox(0,0){$\diamond$}}}
\put(458,440){\raisebox{-.8pt}{\makebox(0,0){$\diamond$}}}
\put(599,402){\raisebox{-.8pt}{\makebox(0,0){$\diamond$}}}
\put(786,294){\raisebox{-.8pt}{\makebox(0,0){$\diamond$}}}
\put(1026,241){\raisebox{-.8pt}{\makebox(0,0){$\diamond$}}}
\put(1326,174){\raisebox{-.8pt}{\makebox(0,0){$\diamond$}}}
\put(1262,723){\makebox(0,0)[r]{Data}}
\put(1338,723){\makebox(0,0){$+$}}
\put(220,752){\makebox(0,0){$+$}}
\put(228,695){\makebox(0,0){$+$}}
\put(249,628){\makebox(0,0){$+$}}
\put(290,578){\makebox(0,0){$+$}}
\put(357,499){\makebox(0,0){$+$}}
\put(458,427){\makebox(0,0){$+$}}
\put(599,369){\makebox(0,0){$+$}}
\put(786,303){\makebox(0,0){$+$}}
\put(1026,267){\makebox(0,0){$+$}}
\put(1326,208){\makebox(0,0){$+$}}
\sbox{\plotpoint}{\rule[-0.400pt]{0.800pt}{0.800pt}}%
\put(1262,678){\makebox(0,0)[r]{Fit to stretched exp}}
\put(1284.0,678.0){\rule[-0.400pt]{26.017pt}{0.800pt}}
\put(220,753){\usebox{\plotpoint}}
\multiput(221.40,729.23)(0.512,-3.730){15}{\rule{0.123pt}{5.727pt}}
\multiput(218.34,741.11)(11.000,-64.113){2}{\rule{0.800pt}{2.864pt}}
\multiput(232.40,665.91)(0.512,-1.636){15}{\rule{0.123pt}{2.673pt}}
\multiput(229.34,671.45)(11.000,-28.453){2}{\rule{0.800pt}{1.336pt}}
\multiput(243.41,635.53)(0.511,-1.033){17}{\rule{0.123pt}{1.800pt}}
\multiput(240.34,639.26)(12.000,-20.264){2}{\rule{0.800pt}{0.900pt}}
\multiput(255.40,612.43)(0.512,-0.888){15}{\rule{0.123pt}{1.582pt}}
\multiput(252.34,615.72)(11.000,-15.717){2}{\rule{0.800pt}{0.791pt}}
\multiput(266.40,594.04)(0.512,-0.788){15}{\rule{0.123pt}{1.436pt}}
\multiput(263.34,597.02)(11.000,-14.019){2}{\rule{0.800pt}{0.718pt}}
\multiput(277.40,577.94)(0.512,-0.639){15}{\rule{0.123pt}{1.218pt}}
\multiput(274.34,580.47)(11.000,-11.472){2}{\rule{0.800pt}{0.609pt}}
\multiput(288.40,564.25)(0.512,-0.589){15}{\rule{0.123pt}{1.145pt}}
\multiput(285.34,566.62)(11.000,-10.623){2}{\rule{0.800pt}{0.573pt}}
\multiput(299.40,551.55)(0.512,-0.539){15}{\rule{0.123pt}{1.073pt}}
\multiput(296.34,553.77)(11.000,-9.774){2}{\rule{0.800pt}{0.536pt}}
\multiput(309.00,542.08)(0.539,-0.512){15}{\rule{1.073pt}{0.123pt}}
\multiput(309.00,542.34)(9.774,-11.000){2}{\rule{0.536pt}{0.800pt}}
\multiput(321.00,531.08)(0.543,-0.514){13}{\rule{1.080pt}{0.124pt}}
\multiput(321.00,531.34)(8.758,-10.000){2}{\rule{0.540pt}{0.800pt}}
\multiput(332.00,521.08)(0.611,-0.516){11}{\rule{1.178pt}{0.124pt}}
\multiput(332.00,521.34)(8.555,-9.000){2}{\rule{0.589pt}{0.800pt}}
\multiput(343.00,512.08)(0.611,-0.516){11}{\rule{1.178pt}{0.124pt}}
\multiput(343.00,512.34)(8.555,-9.000){2}{\rule{0.589pt}{0.800pt}}
\multiput(354.00,503.08)(0.700,-0.520){9}{\rule{1.300pt}{0.125pt}}
\multiput(354.00,503.34)(8.302,-8.000){2}{\rule{0.650pt}{0.800pt}}
\multiput(365.00,495.08)(0.774,-0.520){9}{\rule{1.400pt}{0.125pt}}
\multiput(365.00,495.34)(9.094,-8.000){2}{\rule{0.700pt}{0.800pt}}
\multiput(377.00,487.08)(0.700,-0.520){9}{\rule{1.300pt}{0.125pt}}
\multiput(377.00,487.34)(8.302,-8.000){2}{\rule{0.650pt}{0.800pt}}
\multiput(388.00,479.08)(0.825,-0.526){7}{\rule{1.457pt}{0.127pt}}
\multiput(388.00,479.34)(7.976,-7.000){2}{\rule{0.729pt}{0.800pt}}
\multiput(399.00,472.08)(0.825,-0.526){7}{\rule{1.457pt}{0.127pt}}
\multiput(399.00,472.34)(7.976,-7.000){2}{\rule{0.729pt}{0.800pt}}
\multiput(410.00,465.08)(0.825,-0.526){7}{\rule{1.457pt}{0.127pt}}
\multiput(410.00,465.34)(7.976,-7.000){2}{\rule{0.729pt}{0.800pt}}
\multiput(421.00,458.07)(1.020,-0.536){5}{\rule{1.667pt}{0.129pt}}
\multiput(421.00,458.34)(7.541,-6.000){2}{\rule{0.833pt}{0.800pt}}
\multiput(432.00,452.08)(0.913,-0.526){7}{\rule{1.571pt}{0.127pt}}
\multiput(432.00,452.34)(8.738,-7.000){2}{\rule{0.786pt}{0.800pt}}
\multiput(444.00,445.07)(1.020,-0.536){5}{\rule{1.667pt}{0.129pt}}
\multiput(444.00,445.34)(7.541,-6.000){2}{\rule{0.833pt}{0.800pt}}
\multiput(455.00,439.07)(1.020,-0.536){5}{\rule{1.667pt}{0.129pt}}
\multiput(455.00,439.34)(7.541,-6.000){2}{\rule{0.833pt}{0.800pt}}
\multiput(466.00,433.06)(1.432,-0.560){3}{\rule{1.960pt}{0.135pt}}
\multiput(466.00,433.34)(6.932,-5.000){2}{\rule{0.980pt}{0.800pt}}
\multiput(477.00,428.07)(1.020,-0.536){5}{\rule{1.667pt}{0.129pt}}
\multiput(477.00,428.34)(7.541,-6.000){2}{\rule{0.833pt}{0.800pt}}
\multiput(488.00,422.06)(1.432,-0.560){3}{\rule{1.960pt}{0.135pt}}
\multiput(488.00,422.34)(6.932,-5.000){2}{\rule{0.980pt}{0.800pt}}
\multiput(499.00,417.06)(1.600,-0.560){3}{\rule{2.120pt}{0.135pt}}
\multiput(499.00,417.34)(7.600,-5.000){2}{\rule{1.060pt}{0.800pt}}
\multiput(511.00,412.07)(1.020,-0.536){5}{\rule{1.667pt}{0.129pt}}
\multiput(511.00,412.34)(7.541,-6.000){2}{\rule{0.833pt}{0.800pt}}
\multiput(522.00,406.06)(1.432,-0.560){3}{\rule{1.960pt}{0.135pt}}
\multiput(522.00,406.34)(6.932,-5.000){2}{\rule{0.980pt}{0.800pt}}
\put(533,399.34){\rule{2.400pt}{0.800pt}}
\multiput(533.00,401.34)(6.019,-4.000){2}{\rule{1.200pt}{0.800pt}}
\multiput(544.00,397.06)(1.432,-0.560){3}{\rule{1.960pt}{0.135pt}}
\multiput(544.00,397.34)(6.932,-5.000){2}{\rule{0.980pt}{0.800pt}}
\multiput(555.00,392.06)(1.432,-0.560){3}{\rule{1.960pt}{0.135pt}}
\multiput(555.00,392.34)(6.932,-5.000){2}{\rule{0.980pt}{0.800pt}}
\put(566,385.34){\rule{2.600pt}{0.800pt}}
\multiput(566.00,387.34)(6.604,-4.000){2}{\rule{1.300pt}{0.800pt}}
\multiput(578.00,383.06)(1.432,-0.560){3}{\rule{1.960pt}{0.135pt}}
\multiput(578.00,383.34)(6.932,-5.000){2}{\rule{0.980pt}{0.800pt}}
\put(589,376.34){\rule{2.400pt}{0.800pt}}
\multiput(589.00,378.34)(6.019,-4.000){2}{\rule{1.200pt}{0.800pt}}
\multiput(600.00,374.06)(1.432,-0.560){3}{\rule{1.960pt}{0.135pt}}
\multiput(600.00,374.34)(6.932,-5.000){2}{\rule{0.980pt}{0.800pt}}
\put(611,367.34){\rule{2.400pt}{0.800pt}}
\multiput(611.00,369.34)(6.019,-4.000){2}{\rule{1.200pt}{0.800pt}}
\put(622,363.34){\rule{2.400pt}{0.800pt}}
\multiput(622.00,365.34)(6.019,-4.000){2}{\rule{1.200pt}{0.800pt}}
\put(633,359.34){\rule{2.600pt}{0.800pt}}
\multiput(633.00,361.34)(6.604,-4.000){2}{\rule{1.300pt}{0.800pt}}
\put(645,355.34){\rule{2.400pt}{0.800pt}}
\multiput(645.00,357.34)(6.019,-4.000){2}{\rule{1.200pt}{0.800pt}}
\put(656,351.34){\rule{2.400pt}{0.800pt}}
\multiput(656.00,353.34)(6.019,-4.000){2}{\rule{1.200pt}{0.800pt}}
\put(667,347.34){\rule{2.400pt}{0.800pt}}
\multiput(667.00,349.34)(6.019,-4.000){2}{\rule{1.200pt}{0.800pt}}
\put(678,343.84){\rule{2.650pt}{0.800pt}}
\multiput(678.00,345.34)(5.500,-3.000){2}{\rule{1.325pt}{0.800pt}}
\put(689,340.34){\rule{2.600pt}{0.800pt}}
\multiput(689.00,342.34)(6.604,-4.000){2}{\rule{1.300pt}{0.800pt}}
\put(701,336.34){\rule{2.400pt}{0.800pt}}
\multiput(701.00,338.34)(6.019,-4.000){2}{\rule{1.200pt}{0.800pt}}
\put(712,332.84){\rule{2.650pt}{0.800pt}}
\multiput(712.00,334.34)(5.500,-3.000){2}{\rule{1.325pt}{0.800pt}}
\put(723,329.34){\rule{2.400pt}{0.800pt}}
\multiput(723.00,331.34)(6.019,-4.000){2}{\rule{1.200pt}{0.800pt}}
\put(734,325.34){\rule{2.400pt}{0.800pt}}
\multiput(734.00,327.34)(6.019,-4.000){2}{\rule{1.200pt}{0.800pt}}
\put(745,321.84){\rule{2.650pt}{0.800pt}}
\multiput(745.00,323.34)(5.500,-3.000){2}{\rule{1.325pt}{0.800pt}}
\put(756,318.84){\rule{2.891pt}{0.800pt}}
\multiput(756.00,320.34)(6.000,-3.000){2}{\rule{1.445pt}{0.800pt}}
\put(768,315.34){\rule{2.400pt}{0.800pt}}
\multiput(768.00,317.34)(6.019,-4.000){2}{\rule{1.200pt}{0.800pt}}
\put(779,311.84){\rule{2.650pt}{0.800pt}}
\multiput(779.00,313.34)(5.500,-3.000){2}{\rule{1.325pt}{0.800pt}}
\put(790,308.84){\rule{2.650pt}{0.800pt}}
\multiput(790.00,310.34)(5.500,-3.000){2}{\rule{1.325pt}{0.800pt}}
\put(801,305.34){\rule{2.400pt}{0.800pt}}
\multiput(801.00,307.34)(6.019,-4.000){2}{\rule{1.200pt}{0.800pt}}
\put(812,301.84){\rule{2.650pt}{0.800pt}}
\multiput(812.00,303.34)(5.500,-3.000){2}{\rule{1.325pt}{0.800pt}}
\put(823,298.84){\rule{2.891pt}{0.800pt}}
\multiput(823.00,300.34)(6.000,-3.000){2}{\rule{1.445pt}{0.800pt}}
\put(835,295.84){\rule{2.650pt}{0.800pt}}
\multiput(835.00,297.34)(5.500,-3.000){2}{\rule{1.325pt}{0.800pt}}
\put(846,292.84){\rule{2.650pt}{0.800pt}}
\multiput(846.00,294.34)(5.500,-3.000){2}{\rule{1.325pt}{0.800pt}}
\put(857,289.84){\rule{2.650pt}{0.800pt}}
\multiput(857.00,291.34)(5.500,-3.000){2}{\rule{1.325pt}{0.800pt}}
\put(868,286.84){\rule{2.650pt}{0.800pt}}
\multiput(868.00,288.34)(5.500,-3.000){2}{\rule{1.325pt}{0.800pt}}
\put(879,283.84){\rule{2.650pt}{0.800pt}}
\multiput(879.00,285.34)(5.500,-3.000){2}{\rule{1.325pt}{0.800pt}}
\put(890,280.84){\rule{2.891pt}{0.800pt}}
\multiput(890.00,282.34)(6.000,-3.000){2}{\rule{1.445pt}{0.800pt}}
\put(902,277.84){\rule{2.650pt}{0.800pt}}
\multiput(902.00,279.34)(5.500,-3.000){2}{\rule{1.325pt}{0.800pt}}
\put(913,274.84){\rule{2.650pt}{0.800pt}}
\multiput(913.00,276.34)(5.500,-3.000){2}{\rule{1.325pt}{0.800pt}}
\put(924,271.84){\rule{2.650pt}{0.800pt}}
\multiput(924.00,273.34)(5.500,-3.000){2}{\rule{1.325pt}{0.800pt}}
\put(935,268.84){\rule{2.650pt}{0.800pt}}
\multiput(935.00,270.34)(5.500,-3.000){2}{\rule{1.325pt}{0.800pt}}
\put(946,266.34){\rule{2.650pt}{0.800pt}}
\multiput(946.00,267.34)(5.500,-2.000){2}{\rule{1.325pt}{0.800pt}}
\put(957,263.84){\rule{2.891pt}{0.800pt}}
\multiput(957.00,265.34)(6.000,-3.000){2}{\rule{1.445pt}{0.800pt}}
\put(969,260.84){\rule{2.650pt}{0.800pt}}
\multiput(969.00,262.34)(5.500,-3.000){2}{\rule{1.325pt}{0.800pt}}
\put(980,257.84){\rule{2.650pt}{0.800pt}}
\multiput(980.00,259.34)(5.500,-3.000){2}{\rule{1.325pt}{0.800pt}}
\put(991,255.34){\rule{2.650pt}{0.800pt}}
\multiput(991.00,256.34)(5.500,-2.000){2}{\rule{1.325pt}{0.800pt}}
\put(1002,252.84){\rule{2.650pt}{0.800pt}}
\multiput(1002.00,254.34)(5.500,-3.000){2}{\rule{1.325pt}{0.800pt}}
\put(1013,249.84){\rule{2.891pt}{0.800pt}}
\multiput(1013.00,251.34)(6.000,-3.000){2}{\rule{1.445pt}{0.800pt}}
\put(1025,247.34){\rule{2.650pt}{0.800pt}}
\multiput(1025.00,248.34)(5.500,-2.000){2}{\rule{1.325pt}{0.800pt}}
\put(1036,244.84){\rule{2.650pt}{0.800pt}}
\multiput(1036.00,246.34)(5.500,-3.000){2}{\rule{1.325pt}{0.800pt}}
\put(1047,242.34){\rule{2.650pt}{0.800pt}}
\multiput(1047.00,243.34)(5.500,-2.000){2}{\rule{1.325pt}{0.800pt}}
\put(1058,239.84){\rule{2.650pt}{0.800pt}}
\multiput(1058.00,241.34)(5.500,-3.000){2}{\rule{1.325pt}{0.800pt}}
\put(1069,237.34){\rule{2.650pt}{0.800pt}}
\multiput(1069.00,238.34)(5.500,-2.000){2}{\rule{1.325pt}{0.800pt}}
\put(1080,234.84){\rule{2.891pt}{0.800pt}}
\multiput(1080.00,236.34)(6.000,-3.000){2}{\rule{1.445pt}{0.800pt}}
\put(1092,232.34){\rule{2.650pt}{0.800pt}}
\multiput(1092.00,233.34)(5.500,-2.000){2}{\rule{1.325pt}{0.800pt}}
\put(1103,229.84){\rule{2.650pt}{0.800pt}}
\multiput(1103.00,231.34)(5.500,-3.000){2}{\rule{1.325pt}{0.800pt}}
\put(1114,227.34){\rule{2.650pt}{0.800pt}}
\multiput(1114.00,228.34)(5.500,-2.000){2}{\rule{1.325pt}{0.800pt}}
\put(1125,224.84){\rule{2.650pt}{0.800pt}}
\multiput(1125.00,226.34)(5.500,-3.000){2}{\rule{1.325pt}{0.800pt}}
\put(1136,222.34){\rule{2.650pt}{0.800pt}}
\multiput(1136.00,223.34)(5.500,-2.000){2}{\rule{1.325pt}{0.800pt}}
\put(1147,220.34){\rule{2.891pt}{0.800pt}}
\multiput(1147.00,221.34)(6.000,-2.000){2}{\rule{1.445pt}{0.800pt}}
\put(1159,217.84){\rule{2.650pt}{0.800pt}}
\multiput(1159.00,219.34)(5.500,-3.000){2}{\rule{1.325pt}{0.800pt}}
\put(1170,215.34){\rule{2.650pt}{0.800pt}}
\multiput(1170.00,216.34)(5.500,-2.000){2}{\rule{1.325pt}{0.800pt}}
\put(1181,213.34){\rule{2.650pt}{0.800pt}}
\multiput(1181.00,214.34)(5.500,-2.000){2}{\rule{1.325pt}{0.800pt}}
\put(1192,210.84){\rule{2.650pt}{0.800pt}}
\multiput(1192.00,212.34)(5.500,-3.000){2}{\rule{1.325pt}{0.800pt}}
\put(1203,208.34){\rule{2.650pt}{0.800pt}}
\multiput(1203.00,209.34)(5.500,-2.000){2}{\rule{1.325pt}{0.800pt}}
\put(1214,206.34){\rule{2.891pt}{0.800pt}}
\multiput(1214.00,207.34)(6.000,-2.000){2}{\rule{1.445pt}{0.800pt}}
\put(1226,204.34){\rule{2.650pt}{0.800pt}}
\multiput(1226.00,205.34)(5.500,-2.000){2}{\rule{1.325pt}{0.800pt}}
\put(1237,201.84){\rule{2.650pt}{0.800pt}}
\multiput(1237.00,203.34)(5.500,-3.000){2}{\rule{1.325pt}{0.800pt}}
\put(1248,199.34){\rule{2.650pt}{0.800pt}}
\multiput(1248.00,200.34)(5.500,-2.000){2}{\rule{1.325pt}{0.800pt}}
\put(1259,197.34){\rule{2.650pt}{0.800pt}}
\multiput(1259.00,198.34)(5.500,-2.000){2}{\rule{1.325pt}{0.800pt}}
\put(1270,195.34){\rule{2.650pt}{0.800pt}}
\multiput(1270.00,196.34)(5.500,-2.000){2}{\rule{1.325pt}{0.800pt}}
\put(1281,193.34){\rule{2.891pt}{0.800pt}}
\multiput(1281.00,194.34)(6.000,-2.000){2}{\rule{1.445pt}{0.800pt}}
\put(1293,190.84){\rule{2.650pt}{0.800pt}}
\multiput(1293.00,192.34)(5.500,-3.000){2}{\rule{1.325pt}{0.800pt}}
\put(1304,188.34){\rule{2.650pt}{0.800pt}}
\multiput(1304.00,189.34)(5.500,-2.000){2}{\rule{1.325pt}{0.800pt}}
\put(1315,186.34){\rule{2.650pt}{0.800pt}}
\multiput(1315.00,187.34)(5.500,-2.000){2}{\rule{1.325pt}{0.800pt}}
\end{picture}

%% file: fig5.tex
% GNUPLOT: LaTeX picture
\setlength{\unitlength}{0.240900pt}
\ifx\plotpoint\undefined\newsavebox{\plotpoint}\fi
\sbox{\plotpoint}{\rule[-0.200pt]{0.400pt}{0.400pt}}%
\begin{picture}(1500,900)(0,0)
\font\gnuplot=cmr12 at 12pt
\gnuplot
\sbox{\plotpoint}{\rule[-0.200pt]{0.400pt}{0.400pt}}%
\put(175.0,134.0){\rule[-0.200pt]{4.818pt}{0.400pt}}
\put(153,134){\makebox(0,0)[r]{0}}
\put(1416.0,134.0){\rule[-0.200pt]{4.818pt}{0.400pt}}
\put(175.0,303.0){\rule[-0.200pt]{4.818pt}{0.400pt}}
\put(153,303){\makebox(0,0)[r]{5}}
\put(1416.0,303.0){\rule[-0.200pt]{4.818pt}{0.400pt}}
\put(175.0,472.0){\rule[-0.200pt]{4.818pt}{0.400pt}}
\put(153,472){\makebox(0,0)[r]{10}}
\put(1416.0,472.0){\rule[-0.200pt]{4.818pt}{0.400pt}}
\put(175.0,641.0){\rule[-0.200pt]{4.818pt}{0.400pt}}
\put(153,641){\makebox(0,0)[r]{15}}
\put(1416.0,641.0){\rule[-0.200pt]{4.818pt}{0.400pt}}
\put(175.0,810.0){\rule[-0.200pt]{4.818pt}{0.400pt}}
\put(153,810){\makebox(0,0)[r]{20}}
\put(1416.0,810.0){\rule[-0.200pt]{4.818pt}{0.400pt}}
\put(175.0,134.0){\rule[-0.200pt]{0.400pt}{4.818pt}}
\put(175,89){\makebox(0,0){10}}
\put(175.0,790.0){\rule[-0.200pt]{0.400pt}{4.818pt}}
\put(315.0,134.0){\rule[-0.200pt]{0.400pt}{4.818pt}}
\put(315,89){\makebox(0,0){20}}
\put(315.0,790.0){\rule[-0.200pt]{0.400pt}{4.818pt}}
\put(455.0,134.0){\rule[-0.200pt]{0.400pt}{4.818pt}}
\put(455,89){\makebox(0,0){30}}
\put(455.0,790.0){\rule[-0.200pt]{0.400pt}{4.818pt}}
\put(595.0,134.0){\rule[-0.200pt]{0.400pt}{4.818pt}}
\put(595,89){\makebox(0,0){40}}
\put(595.0,790.0){\rule[-0.200pt]{0.400pt}{4.818pt}}
\put(735.0,134.0){\rule[-0.200pt]{0.400pt}{4.818pt}}
\put(735,89){\makebox(0,0){50}}
\put(735.0,790.0){\rule[-0.200pt]{0.400pt}{4.818pt}}
\put(876.0,134.0){\rule[-0.200pt]{0.400pt}{4.818pt}}
\put(876,89){\makebox(0,0){60}}
\put(876.0,790.0){\rule[-0.200pt]{0.400pt}{4.818pt}}
\put(1016.0,134.0){\rule[-0.200pt]{0.400pt}{4.818pt}}
\put(1016,89){\makebox(0,0){70}}
\put(1016.0,790.0){\rule[-0.200pt]{0.400pt}{4.818pt}}
\put(1156.0,134.0){\rule[-0.200pt]{0.400pt}{4.818pt}}
\put(1156,89){\makebox(0,0){80}}
\put(1156.0,790.0){\rule[-0.200pt]{0.400pt}{4.818pt}}
\put(1296.0,134.0){\rule[-0.200pt]{0.400pt}{4.818pt}}
\put(1296,89){\makebox(0,0){90}}
\put(1296.0,790.0){\rule[-0.200pt]{0.400pt}{4.818pt}}
\put(1436.0,134.0){\rule[-0.200pt]{0.400pt}{4.818pt}}
\put(1436,89){\makebox(0,0){100}}
\put(1436.0,790.0){\rule[-0.200pt]{0.400pt}{4.818pt}}
\put(175.0,134.0){\rule[-0.200pt]{303.775pt}{0.400pt}}
\put(1436.0,134.0){\rule[-0.200pt]{0.400pt}{162.848pt}}
\put(175.0,810.0){\rule[-0.200pt]{303.775pt}{0.400pt}}
\put(45,472){\makebox(0,0){$\langle x\rangle$}}
\put(45,428){\makebox(0,0){(arb. units)}}
\put(805,44){\makebox(0,0){$v_0$ (arb. units)}}
%\put(805,855){\makebox(0,0){Mean stopping distance vs. initial speed}}
\put(175.0,134.0){\rule[-0.200pt]{0.400pt}{162.848pt}}
\put(593,768){\makebox(0,0)[r]{Data}}
\put(669,768){\raisebox{-.8pt}{\makebox(0,0){$\diamond$}}}
\put(175,198){\raisebox{-.8pt}{\makebox(0,0){$\diamond$}}}
\put(315,266){\raisebox{-.8pt}{\makebox(0,0){$\diamond$}}}
\put(455,333){\raisebox{-.8pt}{\makebox(0,0){$\diamond$}}}
\put(595,401){\raisebox{-.8pt}{\makebox(0,0){$\diamond$}}}
\put(735,468){\raisebox{-.8pt}{\makebox(0,0){$\diamond$}}}
\put(876,536){\raisebox{-.8pt}{\makebox(0,0){$\diamond$}}}
\put(1016,604){\raisebox{-.8pt}{\makebox(0,0){$\diamond$}}}
\put(1156,671){\raisebox{-.8pt}{\makebox(0,0){$\diamond$}}}
\put(1296,739){\raisebox{-.8pt}{\makebox(0,0){$\diamond$}}}
\put(1436,806){\raisebox{-.8pt}{\makebox(0,0){$\diamond$}}}
\put(593,723){\makebox(0,0)[r]{Linear fit}}
\multiput(615,723)(20.756,0.000){6}{\usebox{\plotpoint}}
\put(723,723){\usebox{\plotpoint}}
\put(175,201){\usebox{\plotpoint}}
\put(175.00,201.00){\usebox{\plotpoint}}
\put(193.76,209.88){\usebox{\plotpoint}}
\put(212.51,218.77){\usebox{\plotpoint}}
\put(231.19,227.80){\usebox{\plotpoint}}
\put(249.63,237.32){\usebox{\plotpoint}}
\put(268.46,246.06){\usebox{\plotpoint}}
\put(287.30,254.75){\usebox{\plotpoint}}
\put(305.97,263.83){\usebox{\plotpoint}}
\put(324.81,272.53){\usebox{\plotpoint}}
\put(343.62,281.31){\usebox{\plotpoint}}
\put(362.04,290.87){\usebox{\plotpoint}}
\put(380.76,299.81){\usebox{\plotpoint}}
\put(399.49,308.75){\usebox{\plotpoint}}
\put(418.27,317.58){\usebox{\plotpoint}}
\put(437.01,326.50){\usebox{\plotpoint}}
\put(455.78,335.36){\usebox{\plotpoint}}
\put(474.62,344.06){\usebox{\plotpoint}}
\put(493.28,353.15){\usebox{\plotpoint}}
\put(511.72,362.64){\usebox{\plotpoint}}
\put(530.57,371.34){\usebox{\plotpoint}}
\put(549.23,380.42){\usebox{\plotpoint}}
\put(568.08,389.11){\usebox{\plotpoint}}
\put(586.86,397.93){\usebox{\plotpoint}}
\put(605.59,406.89){\usebox{\plotpoint}}
\put(624.33,415.79){\usebox{\plotpoint}}
\put(642.74,425.37){\usebox{\plotpoint}}
\put(661.53,434.17){\usebox{\plotpoint}}
\put(680.26,443.13){\usebox{\plotpoint}}
\put(699.04,451.94){\usebox{\plotpoint}}
\put(717.89,460.64){\usebox{\plotpoint}}
\put(736.55,469.72){\usebox{\plotpoint}}
\put(755.40,478.41){\usebox{\plotpoint}}
\put(773.84,487.91){\usebox{\plotpoint}}
\put(792.50,497.00){\usebox{\plotpoint}}
\put(811.35,505.70){\usebox{\plotpoint}}
\put(830.11,514.56){\usebox{\plotpoint}}
\put(848.85,523.47){\usebox{\plotpoint}}
\put(867.70,532.17){\usebox{\plotpoint}}
\put(886.39,541.19){\usebox{\plotpoint}}
\put(904.80,550.75){\usebox{\plotpoint}}
\put(923.65,559.45){\usebox{\plotpoint}}
\put(942.31,568.53){\usebox{\plotpoint}}
\put(961.16,577.23){\usebox{\plotpoint}}
\put(979.82,586.30){\usebox{\plotpoint}}
\put(998.66,595.00){\usebox{\plotpoint}}
\put(1017.44,603.84){\usebox{\plotpoint}}
\put(1035.74,613.57){\usebox{\plotpoint}}
\put(1054.59,622.27){\usebox{\plotpoint}}
\put(1073.34,631.17){\usebox{\plotpoint}}
\put(1092.10,640.04){\usebox{\plotpoint}}
\put(1110.94,648.74){\usebox{\plotpoint}}
\put(1129.61,657.80){\usebox{\plotpoint}}
\put(1148.45,666.51){\usebox{\plotpoint}}
\put(1166.95,675.90){\usebox{\plotpoint}}
\put(1185.55,685.10){\usebox{\plotpoint}}
\put(1204.40,693.80){\usebox{\plotpoint}}
\put(1223.06,702.87){\usebox{\plotpoint}}
\put(1241.91,711.57){\usebox{\plotpoint}}
\put(1260.71,720.35){\usebox{\plotpoint}}
\put(1279.41,729.35){\usebox{\plotpoint}}
\put(1297.85,738.86){\usebox{\plotpoint}}
\put(1316.58,747.79){\usebox{\plotpoint}}
\put(1335.36,756.63){\usebox{\plotpoint}}
\put(1354.21,765.33){\usebox{\plotpoint}}
\put(1372.87,774.40){\usebox{\plotpoint}}
\put(1391.72,783.10){\usebox{\plotpoint}}
\put(1410.56,791.80){\usebox{\plotpoint}}
\put(1428.79,801.67){\usebox{\plotpoint}}
\put(1436,805){\usebox{\plotpoint}}
\sbox{\plotpoint}{\rule[-0.400pt]{0.800pt}{0.800pt}}%
\put(593,678){\makebox(0,0)[r]{Stretched exp fit}}
\put(615.0,678.0){\rule[-0.400pt]{26.017pt}{0.800pt}}
\put(175,198){\usebox{\plotpoint}}
\multiput(175.00,199.39)(1.244,0.536){5}{\rule{1.933pt}{0.129pt}}
\multiput(175.00,196.34)(8.987,6.000){2}{\rule{0.967pt}{0.800pt}}
\multiput(188.00,205.39)(1.132,0.536){5}{\rule{1.800pt}{0.129pt}}
\multiput(188.00,202.34)(8.264,6.000){2}{\rule{0.900pt}{0.800pt}}
\multiput(200.00,211.39)(1.244,0.536){5}{\rule{1.933pt}{0.129pt}}
\multiput(200.00,208.34)(8.987,6.000){2}{\rule{0.967pt}{0.800pt}}
\multiput(213.00,217.39)(1.244,0.536){5}{\rule{1.933pt}{0.129pt}}
\multiput(213.00,214.34)(8.987,6.000){2}{\rule{0.967pt}{0.800pt}}
\multiput(226.00,223.40)(1.000,0.526){7}{\rule{1.686pt}{0.127pt}}
\multiput(226.00,220.34)(9.501,7.000){2}{\rule{0.843pt}{0.800pt}}
\multiput(239.00,230.39)(1.132,0.536){5}{\rule{1.800pt}{0.129pt}}
\multiput(239.00,227.34)(8.264,6.000){2}{\rule{0.900pt}{0.800pt}}
\multiput(251.00,236.39)(1.244,0.536){5}{\rule{1.933pt}{0.129pt}}
\multiput(251.00,233.34)(8.987,6.000){2}{\rule{0.967pt}{0.800pt}}
\multiput(264.00,242.39)(1.244,0.536){5}{\rule{1.933pt}{0.129pt}}
\multiput(264.00,239.34)(8.987,6.000){2}{\rule{0.967pt}{0.800pt}}
\multiput(277.00,248.39)(1.244,0.536){5}{\rule{1.933pt}{0.129pt}}
\multiput(277.00,245.34)(8.987,6.000){2}{\rule{0.967pt}{0.800pt}}
\multiput(290.00,254.39)(1.132,0.536){5}{\rule{1.800pt}{0.129pt}}
\multiput(290.00,251.34)(8.264,6.000){2}{\rule{0.900pt}{0.800pt}}
\multiput(302.00,260.39)(1.244,0.536){5}{\rule{1.933pt}{0.129pt}}
\multiput(302.00,257.34)(8.987,6.000){2}{\rule{0.967pt}{0.800pt}}
\multiput(315.00,266.40)(1.000,0.526){7}{\rule{1.686pt}{0.127pt}}
\multiput(315.00,263.34)(9.501,7.000){2}{\rule{0.843pt}{0.800pt}}
\multiput(328.00,273.39)(1.244,0.536){5}{\rule{1.933pt}{0.129pt}}
\multiput(328.00,270.34)(8.987,6.000){2}{\rule{0.967pt}{0.800pt}}
\multiput(341.00,279.39)(1.132,0.536){5}{\rule{1.800pt}{0.129pt}}
\multiput(341.00,276.34)(8.264,6.000){2}{\rule{0.900pt}{0.800pt}}
\multiput(353.00,285.39)(1.244,0.536){5}{\rule{1.933pt}{0.129pt}}
\multiput(353.00,282.34)(8.987,6.000){2}{\rule{0.967pt}{0.800pt}}
\multiput(366.00,291.39)(1.244,0.536){5}{\rule{1.933pt}{0.129pt}}
\multiput(366.00,288.34)(8.987,6.000){2}{\rule{0.967pt}{0.800pt}}
\multiput(379.00,297.39)(1.244,0.536){5}{\rule{1.933pt}{0.129pt}}
\multiput(379.00,294.34)(8.987,6.000){2}{\rule{0.967pt}{0.800pt}}
\multiput(392.00,303.39)(1.132,0.536){5}{\rule{1.800pt}{0.129pt}}
\multiput(392.00,300.34)(8.264,6.000){2}{\rule{0.900pt}{0.800pt}}
\multiput(404.00,309.40)(1.000,0.526){7}{\rule{1.686pt}{0.127pt}}
\multiput(404.00,306.34)(9.501,7.000){2}{\rule{0.843pt}{0.800pt}}
\multiput(417.00,316.39)(1.244,0.536){5}{\rule{1.933pt}{0.129pt}}
\multiput(417.00,313.34)(8.987,6.000){2}{\rule{0.967pt}{0.800pt}}
\multiput(430.00,322.39)(1.132,0.536){5}{\rule{1.800pt}{0.129pt}}
\multiput(430.00,319.34)(8.264,6.000){2}{\rule{0.900pt}{0.800pt}}
\multiput(442.00,328.39)(1.244,0.536){5}{\rule{1.933pt}{0.129pt}}
\multiput(442.00,325.34)(8.987,6.000){2}{\rule{0.967pt}{0.800pt}}
\multiput(455.00,334.39)(1.244,0.536){5}{\rule{1.933pt}{0.129pt}}
\multiput(455.00,331.34)(8.987,6.000){2}{\rule{0.967pt}{0.800pt}}
\multiput(468.00,340.39)(1.244,0.536){5}{\rule{1.933pt}{0.129pt}}
\multiput(468.00,337.34)(8.987,6.000){2}{\rule{0.967pt}{0.800pt}}
\multiput(481.00,346.39)(1.132,0.536){5}{\rule{1.800pt}{0.129pt}}
\multiput(481.00,343.34)(8.264,6.000){2}{\rule{0.900pt}{0.800pt}}
\multiput(493.00,352.40)(1.000,0.526){7}{\rule{1.686pt}{0.127pt}}
\multiput(493.00,349.34)(9.501,7.000){2}{\rule{0.843pt}{0.800pt}}
\multiput(506.00,359.39)(1.244,0.536){5}{\rule{1.933pt}{0.129pt}}
\multiput(506.00,356.34)(8.987,6.000){2}{\rule{0.967pt}{0.800pt}}
\multiput(519.00,365.39)(1.244,0.536){5}{\rule{1.933pt}{0.129pt}}
\multiput(519.00,362.34)(8.987,6.000){2}{\rule{0.967pt}{0.800pt}}
\multiput(532.00,371.39)(1.132,0.536){5}{\rule{1.800pt}{0.129pt}}
\multiput(532.00,368.34)(8.264,6.000){2}{\rule{0.900pt}{0.800pt}}
\multiput(544.00,377.39)(1.244,0.536){5}{\rule{1.933pt}{0.129pt}}
\multiput(544.00,374.34)(8.987,6.000){2}{\rule{0.967pt}{0.800pt}}
\multiput(557.00,383.39)(1.244,0.536){5}{\rule{1.933pt}{0.129pt}}
\multiput(557.00,380.34)(8.987,6.000){2}{\rule{0.967pt}{0.800pt}}
\multiput(570.00,389.39)(1.244,0.536){5}{\rule{1.933pt}{0.129pt}}
\multiput(570.00,386.34)(8.987,6.000){2}{\rule{0.967pt}{0.800pt}}
\multiput(583.00,395.40)(0.913,0.526){7}{\rule{1.571pt}{0.127pt}}
\multiput(583.00,392.34)(8.738,7.000){2}{\rule{0.786pt}{0.800pt}}
\multiput(595.00,402.39)(1.244,0.536){5}{\rule{1.933pt}{0.129pt}}
\multiput(595.00,399.34)(8.987,6.000){2}{\rule{0.967pt}{0.800pt}}
\multiput(608.00,408.39)(1.244,0.536){5}{\rule{1.933pt}{0.129pt}}
\multiput(608.00,405.34)(8.987,6.000){2}{\rule{0.967pt}{0.800pt}}
\multiput(621.00,414.39)(1.244,0.536){5}{\rule{1.933pt}{0.129pt}}
\multiput(621.00,411.34)(8.987,6.000){2}{\rule{0.967pt}{0.800pt}}
\multiput(634.00,420.39)(1.132,0.536){5}{\rule{1.800pt}{0.129pt}}
\multiput(634.00,417.34)(8.264,6.000){2}{\rule{0.900pt}{0.800pt}}
\multiput(646.00,426.39)(1.244,0.536){5}{\rule{1.933pt}{0.129pt}}
\multiput(646.00,423.34)(8.987,6.000){2}{\rule{0.967pt}{0.800pt}}
\multiput(659.00,432.40)(1.000,0.526){7}{\rule{1.686pt}{0.127pt}}
\multiput(659.00,429.34)(9.501,7.000){2}{\rule{0.843pt}{0.800pt}}
\multiput(672.00,439.39)(1.132,0.536){5}{\rule{1.800pt}{0.129pt}}
\multiput(672.00,436.34)(8.264,6.000){2}{\rule{0.900pt}{0.800pt}}
\multiput(684.00,445.39)(1.244,0.536){5}{\rule{1.933pt}{0.129pt}}
\multiput(684.00,442.34)(8.987,6.000){2}{\rule{0.967pt}{0.800pt}}
\multiput(697.00,451.39)(1.244,0.536){5}{\rule{1.933pt}{0.129pt}}
\multiput(697.00,448.34)(8.987,6.000){2}{\rule{0.967pt}{0.800pt}}
\multiput(710.00,457.39)(1.244,0.536){5}{\rule{1.933pt}{0.129pt}}
\multiput(710.00,454.34)(8.987,6.000){2}{\rule{0.967pt}{0.800pt}}
\multiput(723.00,463.39)(1.132,0.536){5}{\rule{1.800pt}{0.129pt}}
\multiput(723.00,460.34)(8.264,6.000){2}{\rule{0.900pt}{0.800pt}}
\multiput(735.00,469.39)(1.244,0.536){5}{\rule{1.933pt}{0.129pt}}
\multiput(735.00,466.34)(8.987,6.000){2}{\rule{0.967pt}{0.800pt}}
\multiput(748.00,475.40)(1.000,0.526){7}{\rule{1.686pt}{0.127pt}}
\multiput(748.00,472.34)(9.501,7.000){2}{\rule{0.843pt}{0.800pt}}
\multiput(761.00,482.39)(1.244,0.536){5}{\rule{1.933pt}{0.129pt}}
\multiput(761.00,479.34)(8.987,6.000){2}{\rule{0.967pt}{0.800pt}}
\multiput(774.00,488.39)(1.132,0.536){5}{\rule{1.800pt}{0.129pt}}
\multiput(774.00,485.34)(8.264,6.000){2}{\rule{0.900pt}{0.800pt}}
\multiput(786.00,494.39)(1.244,0.536){5}{\rule{1.933pt}{0.129pt}}
\multiput(786.00,491.34)(8.987,6.000){2}{\rule{0.967pt}{0.800pt}}
\multiput(799.00,500.39)(1.244,0.536){5}{\rule{1.933pt}{0.129pt}}
\multiput(799.00,497.34)(8.987,6.000){2}{\rule{0.967pt}{0.800pt}}
\multiput(812.00,506.39)(1.244,0.536){5}{\rule{1.933pt}{0.129pt}}
\multiput(812.00,503.34)(8.987,6.000){2}{\rule{0.967pt}{0.800pt}}
\multiput(825.00,512.39)(1.132,0.536){5}{\rule{1.800pt}{0.129pt}}
\multiput(825.00,509.34)(8.264,6.000){2}{\rule{0.900pt}{0.800pt}}
\multiput(837.00,518.40)(1.000,0.526){7}{\rule{1.686pt}{0.127pt}}
\multiput(837.00,515.34)(9.501,7.000){2}{\rule{0.843pt}{0.800pt}}
\multiput(850.00,525.39)(1.244,0.536){5}{\rule{1.933pt}{0.129pt}}
\multiput(850.00,522.34)(8.987,6.000){2}{\rule{0.967pt}{0.800pt}}
\multiput(863.00,531.39)(1.244,0.536){5}{\rule{1.933pt}{0.129pt}}
\multiput(863.00,528.34)(8.987,6.000){2}{\rule{0.967pt}{0.800pt}}
\multiput(876.00,537.39)(1.132,0.536){5}{\rule{1.800pt}{0.129pt}}
\multiput(876.00,534.34)(8.264,6.000){2}{\rule{0.900pt}{0.800pt}}
\multiput(888.00,543.39)(1.244,0.536){5}{\rule{1.933pt}{0.129pt}}
\multiput(888.00,540.34)(8.987,6.000){2}{\rule{0.967pt}{0.800pt}}
\multiput(901.00,549.39)(1.244,0.536){5}{\rule{1.933pt}{0.129pt}}
\multiput(901.00,546.34)(8.987,6.000){2}{\rule{0.967pt}{0.800pt}}
\multiput(914.00,555.39)(1.244,0.536){5}{\rule{1.933pt}{0.129pt}}
\multiput(914.00,552.34)(8.987,6.000){2}{\rule{0.967pt}{0.800pt}}
\multiput(927.00,561.40)(0.913,0.526){7}{\rule{1.571pt}{0.127pt}}
\multiput(927.00,558.34)(8.738,7.000){2}{\rule{0.786pt}{0.800pt}}
\multiput(939.00,568.39)(1.244,0.536){5}{\rule{1.933pt}{0.129pt}}
\multiput(939.00,565.34)(8.987,6.000){2}{\rule{0.967pt}{0.800pt}}
\multiput(952.00,574.39)(1.244,0.536){5}{\rule{1.933pt}{0.129pt}}
\multiput(952.00,571.34)(8.987,6.000){2}{\rule{0.967pt}{0.800pt}}
\multiput(965.00,580.39)(1.132,0.536){5}{\rule{1.800pt}{0.129pt}}
\multiput(965.00,577.34)(8.264,6.000){2}{\rule{0.900pt}{0.800pt}}
\multiput(977.00,586.39)(1.244,0.536){5}{\rule{1.933pt}{0.129pt}}
\multiput(977.00,583.34)(8.987,6.000){2}{\rule{0.967pt}{0.800pt}}
\multiput(990.00,592.39)(1.244,0.536){5}{\rule{1.933pt}{0.129pt}}
\multiput(990.00,589.34)(8.987,6.000){2}{\rule{0.967pt}{0.800pt}}
\multiput(1003.00,598.39)(1.244,0.536){5}{\rule{1.933pt}{0.129pt}}
\multiput(1003.00,595.34)(8.987,6.000){2}{\rule{0.967pt}{0.800pt}}
\multiput(1016.00,604.40)(0.913,0.526){7}{\rule{1.571pt}{0.127pt}}
\multiput(1016.00,601.34)(8.738,7.000){2}{\rule{0.786pt}{0.800pt}}
\multiput(1028.00,611.39)(1.244,0.536){5}{\rule{1.933pt}{0.129pt}}
\multiput(1028.00,608.34)(8.987,6.000){2}{\rule{0.967pt}{0.800pt}}
\multiput(1041.00,617.39)(1.244,0.536){5}{\rule{1.933pt}{0.129pt}}
\multiput(1041.00,614.34)(8.987,6.000){2}{\rule{0.967pt}{0.800pt}}
\multiput(1054.00,623.39)(1.244,0.536){5}{\rule{1.933pt}{0.129pt}}
\multiput(1054.00,620.34)(8.987,6.000){2}{\rule{0.967pt}{0.800pt}}
\multiput(1067.00,629.39)(1.132,0.536){5}{\rule{1.800pt}{0.129pt}}
\multiput(1067.00,626.34)(8.264,6.000){2}{\rule{0.900pt}{0.800pt}}
\multiput(1079.00,635.39)(1.244,0.536){5}{\rule{1.933pt}{0.129pt}}
\multiput(1079.00,632.34)(8.987,6.000){2}{\rule{0.967pt}{0.800pt}}
\multiput(1092.00,641.39)(1.244,0.536){5}{\rule{1.933pt}{0.129pt}}
\multiput(1092.00,638.34)(8.987,6.000){2}{\rule{0.967pt}{0.800pt}}
\multiput(1105.00,647.40)(1.000,0.526){7}{\rule{1.686pt}{0.127pt}}
\multiput(1105.00,644.34)(9.501,7.000){2}{\rule{0.843pt}{0.800pt}}
\multiput(1118.00,654.39)(1.132,0.536){5}{\rule{1.800pt}{0.129pt}}
\multiput(1118.00,651.34)(8.264,6.000){2}{\rule{0.900pt}{0.800pt}}
\multiput(1130.00,660.39)(1.244,0.536){5}{\rule{1.933pt}{0.129pt}}
\multiput(1130.00,657.34)(8.987,6.000){2}{\rule{0.967pt}{0.800pt}}
\multiput(1143.00,666.39)(1.244,0.536){5}{\rule{1.933pt}{0.129pt}}
\multiput(1143.00,663.34)(8.987,6.000){2}{\rule{0.967pt}{0.800pt}}
\multiput(1156.00,672.39)(1.244,0.536){5}{\rule{1.933pt}{0.129pt}}
\multiput(1156.00,669.34)(8.987,6.000){2}{\rule{0.967pt}{0.800pt}}
\multiput(1169.00,678.39)(1.132,0.536){5}{\rule{1.800pt}{0.129pt}}
\multiput(1169.00,675.34)(8.264,6.000){2}{\rule{0.900pt}{0.800pt}}
\multiput(1181.00,684.39)(1.244,0.536){5}{\rule{1.933pt}{0.129pt}}
\multiput(1181.00,681.34)(8.987,6.000){2}{\rule{0.967pt}{0.800pt}}
\multiput(1194.00,690.40)(1.000,0.526){7}{\rule{1.686pt}{0.127pt}}
\multiput(1194.00,687.34)(9.501,7.000){2}{\rule{0.843pt}{0.800pt}}
\multiput(1207.00,697.39)(1.132,0.536){5}{\rule{1.800pt}{0.129pt}}
\multiput(1207.00,694.34)(8.264,6.000){2}{\rule{0.900pt}{0.800pt}}
\multiput(1219.00,703.39)(1.244,0.536){5}{\rule{1.933pt}{0.129pt}}
\multiput(1219.00,700.34)(8.987,6.000){2}{\rule{0.967pt}{0.800pt}}
\multiput(1232.00,709.39)(1.244,0.536){5}{\rule{1.933pt}{0.129pt}}
\multiput(1232.00,706.34)(8.987,6.000){2}{\rule{0.967pt}{0.800pt}}
\multiput(1245.00,715.39)(1.244,0.536){5}{\rule{1.933pt}{0.129pt}}
\multiput(1245.00,712.34)(8.987,6.000){2}{\rule{0.967pt}{0.800pt}}
\multiput(1258.00,721.39)(1.132,0.536){5}{\rule{1.800pt}{0.129pt}}
\multiput(1258.00,718.34)(8.264,6.000){2}{\rule{0.900pt}{0.800pt}}
\multiput(1270.00,727.39)(1.244,0.536){5}{\rule{1.933pt}{0.129pt}}
\multiput(1270.00,724.34)(8.987,6.000){2}{\rule{0.967pt}{0.800pt}}
\multiput(1283.00,733.40)(1.000,0.526){7}{\rule{1.686pt}{0.127pt}}
\multiput(1283.00,730.34)(9.501,7.000){2}{\rule{0.843pt}{0.800pt}}
\multiput(1296.00,740.39)(1.244,0.536){5}{\rule{1.933pt}{0.129pt}}
\multiput(1296.00,737.34)(8.987,6.000){2}{\rule{0.967pt}{0.800pt}}
\multiput(1309.00,746.39)(1.132,0.536){5}{\rule{1.800pt}{0.129pt}}
\multiput(1309.00,743.34)(8.264,6.000){2}{\rule{0.900pt}{0.800pt}}
\multiput(1321.00,752.39)(1.244,0.536){5}{\rule{1.933pt}{0.129pt}}
\multiput(1321.00,749.34)(8.987,6.000){2}{\rule{0.967pt}{0.800pt}}
\multiput(1334.00,758.39)(1.244,0.536){5}{\rule{1.933pt}{0.129pt}}
\multiput(1334.00,755.34)(8.987,6.000){2}{\rule{0.967pt}{0.800pt}}
\multiput(1347.00,764.39)(1.244,0.536){5}{\rule{1.933pt}{0.129pt}}
\multiput(1347.00,761.34)(8.987,6.000){2}{\rule{0.967pt}{0.800pt}}
\multiput(1360.00,770.39)(1.132,0.536){5}{\rule{1.800pt}{0.129pt}}
\multiput(1360.00,767.34)(8.264,6.000){2}{\rule{0.900pt}{0.800pt}}
\multiput(1372.00,776.40)(1.000,0.526){7}{\rule{1.686pt}{0.127pt}}
\multiput(1372.00,773.34)(9.501,7.000){2}{\rule{0.843pt}{0.800pt}}
\multiput(1385.00,783.39)(1.244,0.536){5}{\rule{1.933pt}{0.129pt}}
\multiput(1385.00,780.34)(8.987,6.000){2}{\rule{0.967pt}{0.800pt}}
\multiput(1398.00,789.39)(1.244,0.536){5}{\rule{1.933pt}{0.129pt}}
\multiput(1398.00,786.34)(8.987,6.000){2}{\rule{0.967pt}{0.800pt}}
\multiput(1411.00,795.39)(1.132,0.536){5}{\rule{1.800pt}{0.129pt}}
\multiput(1411.00,792.34)(8.264,6.000){2}{\rule{0.900pt}{0.800pt}}
\multiput(1423.00,801.39)(1.244,0.536){5}{\rule{1.933pt}{0.129pt}}
\multiput(1423.00,798.34)(8.987,6.000){2}{\rule{0.967pt}{0.800pt}}
\sbox{\plotpoint}{\rule[-0.500pt]{1.000pt}{1.000pt}}%
\put(593,633){\makebox(0,0)[r]{Eq.~(13)}}
\multiput(615,633)(20.756,0.000){6}{\usebox{\plotpoint}}
\put(723,633){\usebox{\plotpoint}}
\put(175,198){\usebox{\plotpoint}}
\put(175.00,198.00){\usebox{\plotpoint}}
\put(193.76,206.88){\usebox{\plotpoint}}
\put(212.51,215.77){\usebox{\plotpoint}}
\put(231.35,224.47){\usebox{\plotpoint}}
\put(249.65,234.21){\usebox{\plotpoint}}
\put(268.43,243.04){\usebox{\plotpoint}}
\put(287.28,251.74){\usebox{\plotpoint}}
\put(305.94,260.82){\usebox{\plotpoint}}
\put(324.78,269.52){\usebox{\plotpoint}}
\put(343.19,279.10){\usebox{\plotpoint}}
\put(361.89,288.10){\usebox{\plotpoint}}
\put(380.73,296.80){\usebox{\plotpoint}}
\put(399.46,305.73){\usebox{\plotpoint}}
\put(418.20,314.65){\usebox{\plotpoint}}
\put(436.58,324.29){\usebox{\plotpoint}}
\put(455.34,333.16){\usebox{\plotpoint}}
\put(474.19,341.86){\usebox{\plotpoint}}
\put(492.85,350.93){\usebox{\plotpoint}}
\put(511.53,359.98){\usebox{\plotpoint}}
\put(530.14,369.14){\usebox{\plotpoint}}
\put(548.80,378.22){\usebox{\plotpoint}}
\put(567.65,386.91){\usebox{\plotpoint}}
\put(586.32,395.94){\usebox{\plotpoint}}
\put(604.72,405.49){\usebox{\plotpoint}}
\put(623.57,414.19){\usebox{\plotpoint}}
\put(642.29,423.14){\usebox{\plotpoint}}
\put(661.08,431.96){\usebox{\plotpoint}}
\put(679.54,441.40){\usebox{\plotpoint}}
\put(698.15,450.53){\usebox{\plotpoint}}
\put(717.00,459.23){\usebox{\plotpoint}}
\put(735.66,468.31){\usebox{\plotpoint}}
\put(754.51,477.00){\usebox{\plotpoint}}
\put(772.98,486.45){\usebox{\plotpoint}}
\put(791.61,495.59){\usebox{\plotpoint}}
\put(810.45,504.29){\usebox{\plotpoint}}
\put(829.24,513.12){\usebox{\plotpoint}}
\put(847.96,522.06){\usebox{\plotpoint}}
\put(866.40,531.57){\usebox{\plotpoint}}
\put(885.11,540.56){\usebox{\plotpoint}}
\put(903.91,549.34){\usebox{\plotpoint}}
\put(922.76,558.04){\usebox{\plotpoint}}
\put(941.35,567.26){\usebox{\plotpoint}}
\put(959.86,576.63){\usebox{\plotpoint}}
\put(978.52,585.70){\usebox{\plotpoint}}
\put(997.37,594.40){\usebox{\plotpoint}}
\put(1016.21,603.11){\usebox{\plotpoint}}
\put(1034.67,612.59){\usebox{\plotpoint}}
\put(1053.32,621.68){\usebox{\plotpoint}}
\put(1072.08,630.54){\usebox{\plotpoint}}
\put(1090.83,639.46){\usebox{\plotpoint}}
\put(1109.67,648.16){\usebox{\plotpoint}}
\put(1128.00,657.84){\usebox{\plotpoint}}
\put(1146.75,666.73){\usebox{\plotpoint}}
\put(1165.59,675.43){\usebox{\plotpoint}}
\put(1184.26,684.50){\usebox{\plotpoint}}
\put(1203.10,693.20){\usebox{\plotpoint}}
\put(1221.33,703.08){\usebox{\plotpoint}}
\put(1240.18,711.77){\usebox{\plotpoint}}
\put(1259.01,720.50){\usebox{\plotpoint}}
\put(1277.69,729.55){\usebox{\plotpoint}}
\put(1296.51,738.28){\usebox{\plotpoint}}
\put(1314.88,747.94){\usebox{\plotpoint}}
\put(1333.63,756.83){\usebox{\plotpoint}}
\put(1352.48,765.53){\usebox{\plotpoint}}
\put(1371.16,774.58){\usebox{\plotpoint}}
\put(1389.58,784.11){\usebox{\plotpoint}}
\put(1408.43,792.81){\usebox{\plotpoint}}
\put(1427.09,801.89){\usebox{\plotpoint}}
\put(1436,806){\usebox{\plotpoint}}
\end{picture}

%% file: fig6.tex
% GNUPLOT: LaTeX picture
\setlength{\unitlength}{0.240900pt}
\ifx\plotpoint\undefined\newsavebox{\plotpoint}\fi
\sbox{\plotpoint}{\rule[-0.200pt]{0.400pt}{0.400pt}}%
\begin{picture}(1500,900)(0,0)
\font\gnuplot=cmr12 at 12pt
\gnuplot
\sbox{\plotpoint}{\rule[-0.200pt]{0.400pt}{0.400pt}}%
\put(219.0,134.0){\rule[-0.200pt]{4.818pt}{0.400pt}}
\put(197,134){\makebox(0,0)[r]{0.01}}
\put(1416.0,134.0){\rule[-0.200pt]{4.818pt}{0.400pt}}
\put(219.0,236.0){\rule[-0.200pt]{2.409pt}{0.400pt}}
\put(1426.0,236.0){\rule[-0.200pt]{2.409pt}{0.400pt}}
\put(219.0,295.0){\rule[-0.200pt]{2.409pt}{0.400pt}}
\put(1426.0,295.0){\rule[-0.200pt]{2.409pt}{0.400pt}}
\put(219.0,337.0){\rule[-0.200pt]{2.409pt}{0.400pt}}
\put(1426.0,337.0){\rule[-0.200pt]{2.409pt}{0.400pt}}
\put(219.0,370.0){\rule[-0.200pt]{2.409pt}{0.400pt}}
\put(1426.0,370.0){\rule[-0.200pt]{2.409pt}{0.400pt}}
\put(219.0,397.0){\rule[-0.200pt]{2.409pt}{0.400pt}}
\put(1426.0,397.0){\rule[-0.200pt]{2.409pt}{0.400pt}}
\put(219.0,420.0){\rule[-0.200pt]{2.409pt}{0.400pt}}
\put(1426.0,420.0){\rule[-0.200pt]{2.409pt}{0.400pt}}
\put(219.0,439.0){\rule[-0.200pt]{2.409pt}{0.400pt}}
\put(1426.0,439.0){\rule[-0.200pt]{2.409pt}{0.400pt}}
\put(219.0,457.0){\rule[-0.200pt]{2.409pt}{0.400pt}}
\put(1426.0,457.0){\rule[-0.200pt]{2.409pt}{0.400pt}}
\put(219.0,472.0){\rule[-0.200pt]{4.818pt}{0.400pt}}
\put(197,472){\makebox(0,0)[r]{0.1}}
\put(1416.0,472.0){\rule[-0.200pt]{4.818pt}{0.400pt}}
\put(219.0,574.0){\rule[-0.200pt]{2.409pt}{0.400pt}}
\put(1426.0,574.0){\rule[-0.200pt]{2.409pt}{0.400pt}}
\put(219.0,633.0){\rule[-0.200pt]{2.409pt}{0.400pt}}
\put(1426.0,633.0){\rule[-0.200pt]{2.409pt}{0.400pt}}
\put(219.0,675.0){\rule[-0.200pt]{2.409pt}{0.400pt}}
\put(1426.0,675.0){\rule[-0.200pt]{2.409pt}{0.400pt}}
\put(219.0,708.0){\rule[-0.200pt]{2.409pt}{0.400pt}}
\put(1426.0,708.0){\rule[-0.200pt]{2.409pt}{0.400pt}}
\put(219.0,735.0){\rule[-0.200pt]{2.409pt}{0.400pt}}
\put(1426.0,735.0){\rule[-0.200pt]{2.409pt}{0.400pt}}
\put(219.0,758.0){\rule[-0.200pt]{2.409pt}{0.400pt}}
\put(1426.0,758.0){\rule[-0.200pt]{2.409pt}{0.400pt}}
\put(219.0,777.0){\rule[-0.200pt]{2.409pt}{0.400pt}}
\put(1426.0,777.0){\rule[-0.200pt]{2.409pt}{0.400pt}}
\put(219.0,795.0){\rule[-0.200pt]{2.409pt}{0.400pt}}
\put(1426.0,795.0){\rule[-0.200pt]{2.409pt}{0.400pt}}
\put(219.0,810.0){\rule[-0.200pt]{4.818pt}{0.400pt}}
\put(197,810){\makebox(0,0)[r]{1}}
\put(1416.0,810.0){\rule[-0.200pt]{4.818pt}{0.400pt}}
\put(219.0,134.0){\rule[-0.200pt]{0.400pt}{4.818pt}}
\put(219,89){\makebox(0,0){0.001}}
\put(219.0,790.0){\rule[-0.200pt]{0.400pt}{4.818pt}}
\put(311.0,134.0){\rule[-0.200pt]{0.400pt}{2.409pt}}
\put(311.0,800.0){\rule[-0.200pt]{0.400pt}{2.409pt}}
\put(364.0,134.0){\rule[-0.200pt]{0.400pt}{2.409pt}}
\put(364.0,800.0){\rule[-0.200pt]{0.400pt}{2.409pt}}
\put(402.0,134.0){\rule[-0.200pt]{0.400pt}{2.409pt}}
\put(402.0,800.0){\rule[-0.200pt]{0.400pt}{2.409pt}}
\put(432.0,134.0){\rule[-0.200pt]{0.400pt}{2.409pt}}
\put(432.0,800.0){\rule[-0.200pt]{0.400pt}{2.409pt}}
\put(456.0,134.0){\rule[-0.200pt]{0.400pt}{2.409pt}}
\put(456.0,800.0){\rule[-0.200pt]{0.400pt}{2.409pt}}
\put(476.0,134.0){\rule[-0.200pt]{0.400pt}{2.409pt}}
\put(476.0,800.0){\rule[-0.200pt]{0.400pt}{2.409pt}}
\put(494.0,134.0){\rule[-0.200pt]{0.400pt}{2.409pt}}
\put(494.0,800.0){\rule[-0.200pt]{0.400pt}{2.409pt}}
\put(509.0,134.0){\rule[-0.200pt]{0.400pt}{2.409pt}}
\put(509.0,800.0){\rule[-0.200pt]{0.400pt}{2.409pt}}
\put(523.0,134.0){\rule[-0.200pt]{0.400pt}{4.818pt}}
\put(523,89){\makebox(0,0){0.01}}
\put(523.0,790.0){\rule[-0.200pt]{0.400pt}{4.818pt}}
\put(615.0,134.0){\rule[-0.200pt]{0.400pt}{2.409pt}}
\put(615.0,800.0){\rule[-0.200pt]{0.400pt}{2.409pt}}
\put(668.0,134.0){\rule[-0.200pt]{0.400pt}{2.409pt}}
\put(668.0,800.0){\rule[-0.200pt]{0.400pt}{2.409pt}}
\put(706.0,134.0){\rule[-0.200pt]{0.400pt}{2.409pt}}
\put(706.0,800.0){\rule[-0.200pt]{0.400pt}{2.409pt}}
\put(736.0,134.0){\rule[-0.200pt]{0.400pt}{2.409pt}}
\put(736.0,800.0){\rule[-0.200pt]{0.400pt}{2.409pt}}
\put(760.0,134.0){\rule[-0.200pt]{0.400pt}{2.409pt}}
\put(760.0,800.0){\rule[-0.200pt]{0.400pt}{2.409pt}}
\put(780.0,134.0){\rule[-0.200pt]{0.400pt}{2.409pt}}
\put(780.0,800.0){\rule[-0.200pt]{0.400pt}{2.409pt}}
\put(798.0,134.0){\rule[-0.200pt]{0.400pt}{2.409pt}}
\put(798.0,800.0){\rule[-0.200pt]{0.400pt}{2.409pt}}
\put(814.0,134.0){\rule[-0.200pt]{0.400pt}{2.409pt}}
\put(814.0,800.0){\rule[-0.200pt]{0.400pt}{2.409pt}}
\put(828.0,134.0){\rule[-0.200pt]{0.400pt}{4.818pt}}
\put(828,89){\makebox(0,0){0.1}}
\put(828.0,790.0){\rule[-0.200pt]{0.400pt}{4.818pt}}
\put(919.0,134.0){\rule[-0.200pt]{0.400pt}{2.409pt}}
\put(919.0,800.0){\rule[-0.200pt]{0.400pt}{2.409pt}}
\put(973.0,134.0){\rule[-0.200pt]{0.400pt}{2.409pt}}
\put(973.0,800.0){\rule[-0.200pt]{0.400pt}{2.409pt}}
\put(1011.0,134.0){\rule[-0.200pt]{0.400pt}{2.409pt}}
\put(1011.0,800.0){\rule[-0.200pt]{0.400pt}{2.409pt}}
\put(1040.0,134.0){\rule[-0.200pt]{0.400pt}{2.409pt}}
\put(1040.0,800.0){\rule[-0.200pt]{0.400pt}{2.409pt}}
\put(1064.0,134.0){\rule[-0.200pt]{0.400pt}{2.409pt}}
\put(1064.0,800.0){\rule[-0.200pt]{0.400pt}{2.409pt}}
\put(1085.0,134.0){\rule[-0.200pt]{0.400pt}{2.409pt}}
\put(1085.0,800.0){\rule[-0.200pt]{0.400pt}{2.409pt}}
\put(1102.0,134.0){\rule[-0.200pt]{0.400pt}{2.409pt}}
\put(1102.0,800.0){\rule[-0.200pt]{0.400pt}{2.409pt}}
\put(1118.0,134.0){\rule[-0.200pt]{0.400pt}{2.409pt}}
\put(1118.0,800.0){\rule[-0.200pt]{0.400pt}{2.409pt}}
\put(1132.0,134.0){\rule[-0.200pt]{0.400pt}{4.818pt}}
\put(1132,89){\makebox(0,0){1}}
\put(1132.0,790.0){\rule[-0.200pt]{0.400pt}{4.818pt}}
\put(1223.0,134.0){\rule[-0.200pt]{0.400pt}{2.409pt}}
\put(1223.0,800.0){\rule[-0.200pt]{0.400pt}{2.409pt}}
\put(1277.0,134.0){\rule[-0.200pt]{0.400pt}{2.409pt}}
\put(1277.0,800.0){\rule[-0.200pt]{0.400pt}{2.409pt}}
\put(1315.0,134.0){\rule[-0.200pt]{0.400pt}{2.409pt}}
\put(1315.0,800.0){\rule[-0.200pt]{0.400pt}{2.409pt}}
\put(1344.0,134.0){\rule[-0.200pt]{0.400pt}{2.409pt}}
\put(1344.0,800.0){\rule[-0.200pt]{0.400pt}{2.409pt}}
\put(1369.0,134.0){\rule[-0.200pt]{0.400pt}{2.409pt}}
\put(1369.0,800.0){\rule[-0.200pt]{0.400pt}{2.409pt}}
\put(1389.0,134.0){\rule[-0.200pt]{0.400pt}{2.409pt}}
\put(1389.0,800.0){\rule[-0.200pt]{0.400pt}{2.409pt}}
\put(1407.0,134.0){\rule[-0.200pt]{0.400pt}{2.409pt}}
\put(1407.0,800.0){\rule[-0.200pt]{0.400pt}{2.409pt}}
\put(1422.0,134.0){\rule[-0.200pt]{0.400pt}{2.409pt}}
\put(1422.0,800.0){\rule[-0.200pt]{0.400pt}{2.409pt}}
\put(1436.0,134.0){\rule[-0.200pt]{0.400pt}{4.818pt}}
\put(1436,89){\makebox(0,0){10}}
\put(1436.0,790.0){\rule[-0.200pt]{0.400pt}{4.818pt}}
\put(219.0,134.0){\rule[-0.200pt]{293.175pt}{0.400pt}}
\put(1436.0,134.0){\rule[-0.200pt]{0.400pt}{162.848pt}}
\put(219.0,810.0){\rule[-0.200pt]{293.175pt}{0.400pt}}
\put(45,472){\makebox(0,0){$\Delta x$}}
\put(45,428){\makebox(0,0){(arb. units)}}
\put(827,44){\makebox(0,0){$\sigma$ (arb. units)}}
%\put(827,855){\makebox(0,0){Standard deviation vs. disorder strength}}
\put(219.0,134.0){\rule[-0.200pt]{0.400pt}{162.848pt}}
\put(1262,768){\makebox(0,0)[r]{Data}}
\put(1338,768){\raisebox{-.8pt}{\makebox(0,0){$\diamond$}}}
\put(219,204){\raisebox{-.8pt}{\makebox(0,0){$\diamond$}}}
\put(498,298){\raisebox{-.8pt}{\makebox(0,0){$\diamond$}}}
\put(662,353){\raisebox{-.8pt}{\makebox(0,0){$\diamond$}}}
\put(778,400){\raisebox{-.8pt}{\makebox(0,0){$\diamond$}}}
\put(867,433){\raisebox{-.8pt}{\makebox(0,0){$\diamond$}}}
\put(940,457){\raisebox{-.8pt}{\makebox(0,0){$\diamond$}}}
\put(1001,482){\raisebox{-.8pt}{\makebox(0,0){$\diamond$}}}
\put(1054,500){\raisebox{-.8pt}{\makebox(0,0){$\diamond$}}}
\put(1100,516){\raisebox{-.8pt}{\makebox(0,0){$\diamond$}}}
\put(1141,532){\raisebox{-.8pt}{\makebox(0,0){$\diamond$}}}
\put(1262,723){\makebox(0,0)[r]{Powerlaw fit}}
\multiput(1284,723)(20.756,0.000){6}{\usebox{\plotpoint}}
\put(1392,723){\usebox{\plotpoint}}
\put(219,247){\usebox{\plotpoint}}
\put(219.00,247.00){\usebox{\plotpoint}}
\put(238.79,253.26){\usebox{\plotpoint}}
\put(258.50,259.75){\usebox{\plotpoint}}
\put(278.26,266.09){\usebox{\plotpoint}}
\put(298.05,272.35){\usebox{\plotpoint}}
\put(317.74,278.91){\usebox{\plotpoint}}
\put(337.52,285.17){\usebox{\plotpoint}}
\put(357.29,291.49){\usebox{\plotpoint}}
\put(377.00,298.00){\usebox{\plotpoint}}
\put(396.79,304.26){\usebox{\plotpoint}}
\put(416.57,310.52){\usebox{\plotpoint}}
\put(436.29,316.99){\usebox{\plotpoint}}
\put(456.05,323.35){\usebox{\plotpoint}}
\put(475.83,329.61){\usebox{\plotpoint}}
\put(495.71,335.49){\usebox{\plotpoint}}
\put(515.56,341.52){\usebox{\plotpoint}}
\put(535.34,347.80){\usebox{\plotpoint}}
\put(555.05,354.32){\usebox{\plotpoint}}
\put(574.83,360.61){\usebox{\plotpoint}}
\put(594.61,366.87){\usebox{\plotpoint}}
\put(614.34,373.30){\usebox{\plotpoint}}
\put(634.09,379.70){\usebox{\plotpoint}}
\put(653.84,386.05){\usebox{\plotpoint}}
\put(673.56,392.52){\usebox{\plotpoint}}
\put(693.35,398.78){\usebox{\plotpoint}}
\put(713.14,405.05){\usebox{\plotpoint}}
\put(732.84,411.55){\usebox{\plotpoint}}
\put(752.61,417.87){\usebox{\plotpoint}}
\put(772.40,424.13){\usebox{\plotpoint}}
\put(792.14,430.54){\usebox{\plotpoint}}
\put(811.87,436.96){\usebox{\plotpoint}}
\put(831.64,443.29){\usebox{\plotpoint}}
\put(851.35,449.78){\usebox{\plotpoint}}
\put(871.14,456.05){\usebox{\plotpoint}}
\put(890.92,462.31){\usebox{\plotpoint}}
\put(910.64,468.79){\usebox{\plotpoint}}
\put(930.65,474.22){\usebox{\plotpoint}}
\put(950.44,480.48){\usebox{\plotpoint}}
\put(970.19,486.86){\usebox{\plotpoint}}
\put(989.91,493.30){\usebox{\plotpoint}}
\put(1009.69,499.61){\usebox{\plotpoint}}
\put(1029.39,506.12){\usebox{\plotpoint}}
\put(1049.18,512.39){\usebox{\plotpoint}}
\put(1068.96,518.65){\usebox{\plotpoint}}
\put(1088.69,525.11){\usebox{\plotpoint}}
\put(1108.44,531.48){\usebox{\plotpoint}}
\put(1128.22,537.74){\usebox{\plotpoint}}
\put(1141,542){\usebox{\plotpoint}}
\sbox{\plotpoint}{\rule[-0.400pt]{0.800pt}{0.800pt}}%
\put(1262,678){\makebox(0,0)[r]{Powerlaw fit}}
\put(1284.0,678.0){\rule[-0.400pt]{26.017pt}{0.800pt}}
\put(219,196){\usebox{\plotpoint}}
\put(219,196.34){\rule{2.000pt}{0.800pt}}
\multiput(219.00,194.34)(4.849,4.000){2}{\rule{1.000pt}{0.800pt}}
\put(228,199.84){\rule{2.409pt}{0.800pt}}
\multiput(228.00,198.34)(5.000,3.000){2}{\rule{1.204pt}{0.800pt}}
\put(238,202.84){\rule{2.168pt}{0.800pt}}
\multiput(238.00,201.34)(4.500,3.000){2}{\rule{1.084pt}{0.800pt}}
\put(247,206.34){\rule{2.000pt}{0.800pt}}
\multiput(247.00,204.34)(4.849,4.000){2}{\rule{1.000pt}{0.800pt}}
\put(256,209.84){\rule{2.409pt}{0.800pt}}
\multiput(256.00,208.34)(5.000,3.000){2}{\rule{1.204pt}{0.800pt}}
\put(266,213.34){\rule{2.000pt}{0.800pt}}
\multiput(266.00,211.34)(4.849,4.000){2}{\rule{1.000pt}{0.800pt}}
\put(275,216.84){\rule{2.168pt}{0.800pt}}
\multiput(275.00,215.34)(4.500,3.000){2}{\rule{1.084pt}{0.800pt}}
\put(284,219.84){\rule{2.409pt}{0.800pt}}
\multiput(284.00,218.34)(5.000,3.000){2}{\rule{1.204pt}{0.800pt}}
\put(294,223.34){\rule{2.000pt}{0.800pt}}
\multiput(294.00,221.34)(4.849,4.000){2}{\rule{1.000pt}{0.800pt}}
\put(303,226.84){\rule{2.168pt}{0.800pt}}
\multiput(303.00,225.34)(4.500,3.000){2}{\rule{1.084pt}{0.800pt}}
\put(312,230.34){\rule{2.000pt}{0.800pt}}
\multiput(312.00,228.34)(4.849,4.000){2}{\rule{1.000pt}{0.800pt}}
\put(321,233.84){\rule{2.409pt}{0.800pt}}
\multiput(321.00,232.34)(5.000,3.000){2}{\rule{1.204pt}{0.800pt}}
\put(331,236.84){\rule{2.168pt}{0.800pt}}
\multiput(331.00,235.34)(4.500,3.000){2}{\rule{1.084pt}{0.800pt}}
\put(340,240.34){\rule{2.000pt}{0.800pt}}
\multiput(340.00,238.34)(4.849,4.000){2}{\rule{1.000pt}{0.800pt}}
\put(349,243.84){\rule{2.409pt}{0.800pt}}
\multiput(349.00,242.34)(5.000,3.000){2}{\rule{1.204pt}{0.800pt}}
\put(359,246.84){\rule{2.168pt}{0.800pt}}
\multiput(359.00,245.34)(4.500,3.000){2}{\rule{1.084pt}{0.800pt}}
\put(368,250.34){\rule{2.000pt}{0.800pt}}
\multiput(368.00,248.34)(4.849,4.000){2}{\rule{1.000pt}{0.800pt}}
\put(377,253.84){\rule{2.409pt}{0.800pt}}
\multiput(377.00,252.34)(5.000,3.000){2}{\rule{1.204pt}{0.800pt}}
\put(387,257.34){\rule{2.000pt}{0.800pt}}
\multiput(387.00,255.34)(4.849,4.000){2}{\rule{1.000pt}{0.800pt}}
\put(396,260.84){\rule{2.168pt}{0.800pt}}
\multiput(396.00,259.34)(4.500,3.000){2}{\rule{1.084pt}{0.800pt}}
\put(405,263.84){\rule{2.409pt}{0.800pt}}
\multiput(405.00,262.34)(5.000,3.000){2}{\rule{1.204pt}{0.800pt}}
\put(415,267.34){\rule{2.000pt}{0.800pt}}
\multiput(415.00,265.34)(4.849,4.000){2}{\rule{1.000pt}{0.800pt}}
\put(424,270.84){\rule{2.168pt}{0.800pt}}
\multiput(424.00,269.34)(4.500,3.000){2}{\rule{1.084pt}{0.800pt}}
\put(433,274.34){\rule{2.200pt}{0.800pt}}
\multiput(433.00,272.34)(5.434,4.000){2}{\rule{1.100pt}{0.800pt}}
\put(443,277.84){\rule{2.168pt}{0.800pt}}
\multiput(443.00,276.34)(4.500,3.000){2}{\rule{1.084pt}{0.800pt}}
\put(452,280.84){\rule{2.168pt}{0.800pt}}
\multiput(452.00,279.34)(4.500,3.000){2}{\rule{1.084pt}{0.800pt}}
\put(461,284.34){\rule{2.200pt}{0.800pt}}
\multiput(461.00,282.34)(5.434,4.000){2}{\rule{1.100pt}{0.800pt}}
\put(471,287.84){\rule{2.168pt}{0.800pt}}
\multiput(471.00,286.34)(4.500,3.000){2}{\rule{1.084pt}{0.800pt}}
\put(480,291.34){\rule{2.000pt}{0.800pt}}
\multiput(480.00,289.34)(4.849,4.000){2}{\rule{1.000pt}{0.800pt}}
\put(489,294.84){\rule{2.168pt}{0.800pt}}
\multiput(489.00,293.34)(4.500,3.000){2}{\rule{1.084pt}{0.800pt}}
\put(498,297.84){\rule{2.409pt}{0.800pt}}
\multiput(498.00,296.34)(5.000,3.000){2}{\rule{1.204pt}{0.800pt}}
\put(508,301.34){\rule{2.000pt}{0.800pt}}
\multiput(508.00,299.34)(4.849,4.000){2}{\rule{1.000pt}{0.800pt}}
\put(517,304.84){\rule{2.168pt}{0.800pt}}
\multiput(517.00,303.34)(4.500,3.000){2}{\rule{1.084pt}{0.800pt}}
\put(526,307.84){\rule{2.409pt}{0.800pt}}
\multiput(526.00,306.34)(5.000,3.000){2}{\rule{1.204pt}{0.800pt}}
\put(536,311.34){\rule{2.000pt}{0.800pt}}
\multiput(536.00,309.34)(4.849,4.000){2}{\rule{1.000pt}{0.800pt}}
\put(545,314.84){\rule{2.168pt}{0.800pt}}
\multiput(545.00,313.34)(4.500,3.000){2}{\rule{1.084pt}{0.800pt}}
\put(554,318.34){\rule{2.200pt}{0.800pt}}
\multiput(554.00,316.34)(5.434,4.000){2}{\rule{1.100pt}{0.800pt}}
\put(564,321.84){\rule{2.168pt}{0.800pt}}
\multiput(564.00,320.34)(4.500,3.000){2}{\rule{1.084pt}{0.800pt}}
\put(573,324.84){\rule{2.168pt}{0.800pt}}
\multiput(573.00,323.34)(4.500,3.000){2}{\rule{1.084pt}{0.800pt}}
\put(582,328.34){\rule{2.200pt}{0.800pt}}
\multiput(582.00,326.34)(5.434,4.000){2}{\rule{1.100pt}{0.800pt}}
\put(592,331.84){\rule{2.168pt}{0.800pt}}
\multiput(592.00,330.34)(4.500,3.000){2}{\rule{1.084pt}{0.800pt}}
\put(601,335.34){\rule{2.000pt}{0.800pt}}
\multiput(601.00,333.34)(4.849,4.000){2}{\rule{1.000pt}{0.800pt}}
\put(610,338.84){\rule{2.409pt}{0.800pt}}
\multiput(610.00,337.34)(5.000,3.000){2}{\rule{1.204pt}{0.800pt}}
\put(620,341.84){\rule{2.168pt}{0.800pt}}
\multiput(620.00,340.34)(4.500,3.000){2}{\rule{1.084pt}{0.800pt}}
\put(629,345.34){\rule{2.000pt}{0.800pt}}
\multiput(629.00,343.34)(4.849,4.000){2}{\rule{1.000pt}{0.800pt}}
\put(638,348.84){\rule{2.168pt}{0.800pt}}
\multiput(638.00,347.34)(4.500,3.000){2}{\rule{1.084pt}{0.800pt}}
\put(647,352.34){\rule{2.200pt}{0.800pt}}
\multiput(647.00,350.34)(5.434,4.000){2}{\rule{1.100pt}{0.800pt}}
\put(657,355.84){\rule{2.168pt}{0.800pt}}
\multiput(657.00,354.34)(4.500,3.000){2}{\rule{1.084pt}{0.800pt}}
\put(666,358.84){\rule{2.168pt}{0.800pt}}
\multiput(666.00,357.34)(4.500,3.000){2}{\rule{1.084pt}{0.800pt}}
\put(675,362.34){\rule{2.200pt}{0.800pt}}
\multiput(675.00,360.34)(5.434,4.000){2}{\rule{1.100pt}{0.800pt}}
\put(685,365.84){\rule{2.168pt}{0.800pt}}
\multiput(685.00,364.34)(4.500,3.000){2}{\rule{1.084pt}{0.800pt}}
\put(694,368.84){\rule{2.168pt}{0.800pt}}
\multiput(694.00,367.34)(4.500,3.000){2}{\rule{1.084pt}{0.800pt}}
\put(703,372.34){\rule{2.200pt}{0.800pt}}
\multiput(703.00,370.34)(5.434,4.000){2}{\rule{1.100pt}{0.800pt}}
\put(713,375.84){\rule{2.168pt}{0.800pt}}
\multiput(713.00,374.34)(4.500,3.000){2}{\rule{1.084pt}{0.800pt}}
\put(722,379.34){\rule{2.000pt}{0.800pt}}
\multiput(722.00,377.34)(4.849,4.000){2}{\rule{1.000pt}{0.800pt}}
\put(731,382.84){\rule{2.409pt}{0.800pt}}
\multiput(731.00,381.34)(5.000,3.000){2}{\rule{1.204pt}{0.800pt}}
\put(741,385.84){\rule{2.168pt}{0.800pt}}
\multiput(741.00,384.34)(4.500,3.000){2}{\rule{1.084pt}{0.800pt}}
\put(750,389.34){\rule{2.000pt}{0.800pt}}
\multiput(750.00,387.34)(4.849,4.000){2}{\rule{1.000pt}{0.800pt}}
\put(759,392.84){\rule{2.409pt}{0.800pt}}
\multiput(759.00,391.34)(5.000,3.000){2}{\rule{1.204pt}{0.800pt}}
\put(769,396.34){\rule{2.000pt}{0.800pt}}
\multiput(769.00,394.34)(4.849,4.000){2}{\rule{1.000pt}{0.800pt}}
\put(778,399.84){\rule{2.168pt}{0.800pt}}
\multiput(778.00,398.34)(4.500,3.000){2}{\rule{1.084pt}{0.800pt}}
\put(787,402.84){\rule{2.409pt}{0.800pt}}
\multiput(787.00,401.34)(5.000,3.000){2}{\rule{1.204pt}{0.800pt}}
\put(797,406.34){\rule{2.000pt}{0.800pt}}
\multiput(797.00,404.34)(4.849,4.000){2}{\rule{1.000pt}{0.800pt}}
\put(806,409.84){\rule{2.168pt}{0.800pt}}
\multiput(806.00,408.34)(4.500,3.000){2}{\rule{1.084pt}{0.800pt}}
\put(815,413.34){\rule{2.000pt}{0.800pt}}
\multiput(815.00,411.34)(4.849,4.000){2}{\rule{1.000pt}{0.800pt}}
\put(824,416.84){\rule{2.409pt}{0.800pt}}
\multiput(824.00,415.34)(5.000,3.000){2}{\rule{1.204pt}{0.800pt}}
\put(834,419.84){\rule{2.168pt}{0.800pt}}
\multiput(834.00,418.34)(4.500,3.000){2}{\rule{1.084pt}{0.800pt}}
\put(843,423.34){\rule{2.000pt}{0.800pt}}
\multiput(843.00,421.34)(4.849,4.000){2}{\rule{1.000pt}{0.800pt}}
\put(852,426.84){\rule{2.409pt}{0.800pt}}
\multiput(852.00,425.34)(5.000,3.000){2}{\rule{1.204pt}{0.800pt}}
\put(862,429.84){\rule{2.168pt}{0.800pt}}
\multiput(862.00,428.34)(4.500,3.000){2}{\rule{1.084pt}{0.800pt}}
\put(871,433.34){\rule{2.000pt}{0.800pt}}
\multiput(871.00,431.34)(4.849,4.000){2}{\rule{1.000pt}{0.800pt}}
\put(880,436.84){\rule{2.409pt}{0.800pt}}
\multiput(880.00,435.34)(5.000,3.000){2}{\rule{1.204pt}{0.800pt}}
\put(890,440.34){\rule{2.000pt}{0.800pt}}
\multiput(890.00,438.34)(4.849,4.000){2}{\rule{1.000pt}{0.800pt}}
\put(899,443.84){\rule{2.168pt}{0.800pt}}
\multiput(899.00,442.34)(4.500,3.000){2}{\rule{1.084pt}{0.800pt}}
\put(908,446.84){\rule{2.409pt}{0.800pt}}
\multiput(908.00,445.34)(5.000,3.000){2}{\rule{1.204pt}{0.800pt}}
\put(918,450.34){\rule{2.000pt}{0.800pt}}
\multiput(918.00,448.34)(4.849,4.000){2}{\rule{1.000pt}{0.800pt}}
\put(927,453.84){\rule{2.168pt}{0.800pt}}
\multiput(927.00,452.34)(4.500,3.000){2}{\rule{1.084pt}{0.800pt}}
\put(936,457.34){\rule{2.200pt}{0.800pt}}
\multiput(936.00,455.34)(5.434,4.000){2}{\rule{1.100pt}{0.800pt}}
\put(946,460.84){\rule{2.168pt}{0.800pt}}
\multiput(946.00,459.34)(4.500,3.000){2}{\rule{1.084pt}{0.800pt}}
\put(955,463.84){\rule{2.168pt}{0.800pt}}
\multiput(955.00,462.34)(4.500,3.000){2}{\rule{1.084pt}{0.800pt}}
\put(964,467.34){\rule{2.200pt}{0.800pt}}
\multiput(964.00,465.34)(5.434,4.000){2}{\rule{1.100pt}{0.800pt}}
\put(974,470.84){\rule{2.168pt}{0.800pt}}
\multiput(974.00,469.34)(4.500,3.000){2}{\rule{1.084pt}{0.800pt}}
\put(983,474.34){\rule{2.000pt}{0.800pt}}
\multiput(983.00,472.34)(4.849,4.000){2}{\rule{1.000pt}{0.800pt}}
\put(992,477.84){\rule{2.168pt}{0.800pt}}
\multiput(992.00,476.34)(4.500,3.000){2}{\rule{1.084pt}{0.800pt}}
\put(1001,480.84){\rule{2.409pt}{0.800pt}}
\multiput(1001.00,479.34)(5.000,3.000){2}{\rule{1.204pt}{0.800pt}}
\put(1011,484.34){\rule{2.000pt}{0.800pt}}
\multiput(1011.00,482.34)(4.849,4.000){2}{\rule{1.000pt}{0.800pt}}
\put(1020,487.84){\rule{2.168pt}{0.800pt}}
\multiput(1020.00,486.34)(4.500,3.000){2}{\rule{1.084pt}{0.800pt}}
\put(1029,490.84){\rule{2.409pt}{0.800pt}}
\multiput(1029.00,489.34)(5.000,3.000){2}{\rule{1.204pt}{0.800pt}}
\put(1039,494.34){\rule{2.000pt}{0.800pt}}
\multiput(1039.00,492.34)(4.849,4.000){2}{\rule{1.000pt}{0.800pt}}
\put(1048,497.84){\rule{2.168pt}{0.800pt}}
\multiput(1048.00,496.34)(4.500,3.000){2}{\rule{1.084pt}{0.800pt}}
\put(1057,501.34){\rule{2.200pt}{0.800pt}}
\multiput(1057.00,499.34)(5.434,4.000){2}{\rule{1.100pt}{0.800pt}}
\put(1067,504.84){\rule{2.168pt}{0.800pt}}
\multiput(1067.00,503.34)(4.500,3.000){2}{\rule{1.084pt}{0.800pt}}
\put(1076,507.84){\rule{2.168pt}{0.800pt}}
\multiput(1076.00,506.34)(4.500,3.000){2}{\rule{1.084pt}{0.800pt}}
\put(1085,511.34){\rule{2.200pt}{0.800pt}}
\multiput(1085.00,509.34)(5.434,4.000){2}{\rule{1.100pt}{0.800pt}}
\put(1095,514.84){\rule{2.168pt}{0.800pt}}
\multiput(1095.00,513.34)(4.500,3.000){2}{\rule{1.084pt}{0.800pt}}
\put(1104,518.34){\rule{2.000pt}{0.800pt}}
\multiput(1104.00,516.34)(4.849,4.000){2}{\rule{1.000pt}{0.800pt}}
\put(1113,521.84){\rule{2.409pt}{0.800pt}}
\multiput(1113.00,520.34)(5.000,3.000){2}{\rule{1.204pt}{0.800pt}}
\put(1123,524.84){\rule{2.168pt}{0.800pt}}
\multiput(1123.00,523.34)(4.500,3.000){2}{\rule{1.084pt}{0.800pt}}
\put(1132,528.34){\rule{2.000pt}{0.800pt}}
\multiput(1132.00,526.34)(4.849,4.000){2}{\rule{1.000pt}{0.800pt}}
\sbox{\plotpoint}{\rule[-0.500pt]{1.000pt}{1.000pt}}%
\put(1262,633){\makebox(0,0)[r]{Data}}
\put(1338,633){\makebox(0,0){$+$}}
\put(219,274){\makebox(0,0){$+$}}
\put(498,340){\makebox(0,0){$+$}}
\put(662,387){\makebox(0,0){$+$}}
\put(778,425){\makebox(0,0){$+$}}
\put(867,451){\makebox(0,0){$+$}}
\put(940,477){\makebox(0,0){$+$}}
\put(1001,496){\makebox(0,0){$+$}}
\put(1054,511){\makebox(0,0){$+$}}
\put(1100,531){\makebox(0,0){$+$}}
\put(1141,543){\makebox(0,0){$+$}}
\end{picture}

%% file: fig7.tex
% GNUPLOT: LaTeX picture
\setlength{\unitlength}{0.240900pt}
\ifx\plotpoint\undefined\newsavebox{\plotpoint}\fi
\begin{picture}(1500,900)(0,0)
\font\gnuplot=cmr12 at 12pt
\gnuplot
\sbox{\plotpoint}{\rule[-0.200pt]{0.400pt}{0.400pt}}%
\put(219.0,134.0){\rule[-0.200pt]{4.818pt}{0.400pt}}
\put(197,134){\makebox(0,0)[r]{0.28}}
\put(1416.0,134.0){\rule[-0.200pt]{4.818pt}{0.400pt}}
\put(219.0,247.0){\rule[-0.200pt]{4.818pt}{0.400pt}}
\put(197,247){\makebox(0,0)[r]{0.29}}
\put(1416.0,247.0){\rule[-0.200pt]{4.818pt}{0.400pt}}
\put(219.0,359.0){\rule[-0.200pt]{4.818pt}{0.400pt}}
\put(197,359){\makebox(0,0)[r]{0.3}}
\put(1416.0,359.0){\rule[-0.200pt]{4.818pt}{0.400pt}}
\put(219.0,472.0){\rule[-0.200pt]{4.818pt}{0.400pt}}
\put(197,472){\makebox(0,0)[r]{0.31}}
\put(1416.0,472.0){\rule[-0.200pt]{4.818pt}{0.400pt}}
\put(219.0,585.0){\rule[-0.200pt]{4.818pt}{0.400pt}}
\put(197,585){\makebox(0,0)[r]{0.32}}
\put(1416.0,585.0){\rule[-0.200pt]{4.818pt}{0.400pt}}
\put(219.0,697.0){\rule[-0.200pt]{4.818pt}{0.400pt}}
\put(197,697){\makebox(0,0)[r]{0.33}}
\put(1416.0,697.0){\rule[-0.200pt]{4.818pt}{0.400pt}}
\put(219.0,810.0){\rule[-0.200pt]{4.818pt}{0.400pt}}
\put(197,810){\makebox(0,0)[r]{0.34}}
\put(1416.0,810.0){\rule[-0.200pt]{4.818pt}{0.400pt}}
\put(219.0,134.0){\rule[-0.200pt]{0.400pt}{4.818pt}}
\put(219,89){\makebox(0,0){0.001}}
\put(219.0,790.0){\rule[-0.200pt]{0.400pt}{4.818pt}}
\put(354.0,134.0){\rule[-0.200pt]{0.400pt}{4.818pt}}
\put(354,89){\makebox(0,0){0.002}}
\put(354.0,790.0){\rule[-0.200pt]{0.400pt}{4.818pt}}
\put(489.0,134.0){\rule[-0.200pt]{0.400pt}{4.818pt}}
\put(489,89){\makebox(0,0){0.003}}
\put(489.0,790.0){\rule[-0.200pt]{0.400pt}{4.818pt}}
\put(625.0,134.0){\rule[-0.200pt]{0.400pt}{4.818pt}}
\put(625,89){\makebox(0,0){0.004}}
\put(625.0,790.0){\rule[-0.200pt]{0.400pt}{4.818pt}}
\put(760.0,134.0){\rule[-0.200pt]{0.400pt}{4.818pt}}
\put(760,89){\makebox(0,0){0.005}}
\put(760.0,790.0){\rule[-0.200pt]{0.400pt}{4.818pt}}
\put(895.0,134.0){\rule[-0.200pt]{0.400pt}{4.818pt}}
\put(895,89){\makebox(0,0){0.006}}
\put(895.0,790.0){\rule[-0.200pt]{0.400pt}{4.818pt}}
\put(1030.0,134.0){\rule[-0.200pt]{0.400pt}{4.818pt}}
\put(1030,89){\makebox(0,0){0.007}}
\put(1030.0,790.0){\rule[-0.200pt]{0.400pt}{4.818pt}}
\put(1166.0,134.0){\rule[-0.200pt]{0.400pt}{4.818pt}}
\put(1166,89){\makebox(0,0){0.008}}
\put(1166.0,790.0){\rule[-0.200pt]{0.400pt}{4.818pt}}
\put(1301.0,134.0){\rule[-0.200pt]{0.400pt}{4.818pt}}
\put(1301,89){\makebox(0,0){0.009}}
\put(1301.0,790.0){\rule[-0.200pt]{0.400pt}{4.818pt}}
\put(1436.0,134.0){\rule[-0.200pt]{0.400pt}{4.818pt}}
\put(1436,89){\makebox(0,0){0.01}}
\put(1436.0,790.0){\rule[-0.200pt]{0.400pt}{4.818pt}}
\put(219.0,134.0){\rule[-0.200pt]{293.175pt}{0.400pt}}
\put(1436.0,134.0){\rule[-0.200pt]{0.400pt}{162.848pt}}
\put(219.0,810.0){\rule[-0.200pt]{293.175pt}{0.400pt}}
\put(45,472){\makebox(0,0){$z_1$}}
\put(45,428){\makebox(0,0){(arb. units)}}
\put(827,44){\makebox(0,0){$a_0$ (arb. units)}}
%\put(827,855){\makebox(0,0){Powerlaw exponent vs. lattice constant}}
\put(219.0,134.0){\rule[-0.200pt]{0.400pt}{162.848pt}}
\put(1262,768){\makebox(0,0)[r]{Data}}
\put(1338,768){\raisebox{-.8pt}{\makebox(0,0){$\diamond$}}}
\put(1436,222){\raisebox{-.8pt}{\makebox(0,0){$\diamond$}}}
\put(760,367){\raisebox{-.8pt}{\makebox(0,0){$\diamond$}}}
\put(534,492){\raisebox{-.8pt}{\makebox(0,0){$\diamond$}}}
\put(422,499){\raisebox{-.8pt}{\makebox(0,0){$\diamond$}}}
\put(354,528){\raisebox{-.8pt}{\makebox(0,0){$\diamond$}}}
\put(310,485){\raisebox{-.8pt}{\makebox(0,0){$\diamond$}}}
\put(277,616){\raisebox{-.8pt}{\makebox(0,0){$\diamond$}}}
\put(253,547){\raisebox{-.8pt}{\makebox(0,0){$\diamond$}}}
\put(234,696){\raisebox{-.8pt}{\makebox(0,0){$\diamond$}}}
\put(219,669){\raisebox{-.8pt}{\makebox(0,0){$\diamond$}}}
\put(1284.0,768.0){\rule[-0.200pt]{26.017pt}{0.400pt}}
\put(1284.0,758.0){\rule[-0.200pt]{0.400pt}{4.818pt}}
\put(1392.0,758.0){\rule[-0.200pt]{0.400pt}{4.818pt}}
\put(1436.0,158.0){\rule[-0.200pt]{0.400pt}{30.835pt}}
\put(1426.0,158.0){\rule[-0.200pt]{4.818pt}{0.400pt}}
\put(1426.0,286.0){\rule[-0.200pt]{4.818pt}{0.400pt}}
\put(760.0,337.0){\rule[-0.200pt]{0.400pt}{14.213pt}}
\put(750.0,337.0){\rule[-0.200pt]{4.818pt}{0.400pt}}
\put(750.0,396.0){\rule[-0.200pt]{4.818pt}{0.400pt}}
\put(534.0,454.0){\rule[-0.200pt]{0.400pt}{18.308pt}}
\put(524.0,454.0){\rule[-0.200pt]{4.818pt}{0.400pt}}
\put(524.0,530.0){\rule[-0.200pt]{4.818pt}{0.400pt}}
\put(422.0,479.0){\rule[-0.200pt]{0.400pt}{9.877pt}}
\put(412.0,479.0){\rule[-0.200pt]{4.818pt}{0.400pt}}
\put(412.0,520.0){\rule[-0.200pt]{4.818pt}{0.400pt}}
\put(354.0,494.0){\rule[-0.200pt]{0.400pt}{16.381pt}}
\put(344.0,494.0){\rule[-0.200pt]{4.818pt}{0.400pt}}
\put(344.0,562.0){\rule[-0.200pt]{4.818pt}{0.400pt}}
\put(310.0,443.0){\rule[-0.200pt]{0.400pt}{19.995pt}}
\put(300.0,443.0){\rule[-0.200pt]{4.818pt}{0.400pt}}
\put(300.0,526.0){\rule[-0.200pt]{4.818pt}{0.400pt}}
\put(277.0,584.0){\rule[-0.200pt]{0.400pt}{15.177pt}}
\put(267.0,584.0){\rule[-0.200pt]{4.818pt}{0.400pt}}
\put(267.0,647.0){\rule[-0.200pt]{4.818pt}{0.400pt}}
\put(253.0,517.0){\rule[-0.200pt]{0.400pt}{14.695pt}}
\put(243.0,517.0){\rule[-0.200pt]{4.818pt}{0.400pt}}
\put(243.0,578.0){\rule[-0.200pt]{4.818pt}{0.400pt}}
\put(234.0,659.0){\rule[-0.200pt]{0.400pt}{18.067pt}}
\put(224.0,659.0){\rule[-0.200pt]{4.818pt}{0.400pt}}
\put(224.0,734.0){\rule[-0.200pt]{4.818pt}{0.400pt}}
\put(219.0,645.0){\rule[-0.200pt]{0.400pt}{11.804pt}}
\put(209.0,645.0){\rule[-0.200pt]{4.818pt}{0.400pt}}
\put(209.0,694.0){\rule[-0.200pt]{4.818pt}{0.400pt}}
\put(1262,723){\makebox(0,0)[r]{Linear fit}}
\multiput(1284,723)(20.756,0.000){6}{\usebox{\plotpoint}}
\put(1392,723){\usebox{\plotpoint}}
\put(219,601){\usebox{\plotpoint}}
\put(219.00,601.00){\usebox{\plotpoint}}
\put(238.75,594.62){\usebox{\plotpoint}}
\put(258.14,587.29){\usebox{\plotpoint}}
\put(277.83,580.72){\usebox{\plotpoint}}
\put(297.62,574.46){\usebox{\plotpoint}}
\put(316.98,567.01){\usebox{\plotpoint}}
\put(336.77,560.74){\usebox{\plotpoint}}
\put(356.47,554.24){\usebox{\plotpoint}}
\put(376.24,547.92){\usebox{\plotpoint}}
\put(395.60,540.47){\usebox{\plotpoint}}
\put(415.38,534.19){\usebox{\plotpoint}}
\put(435.08,527.64){\usebox{\plotpoint}}
\put(454.56,520.48){\usebox{\plotpoint}}
\put(474.25,513.92){\usebox{\plotpoint}}
\put(493.97,507.47){\usebox{\plotpoint}}
\put(513.41,500.25){\usebox{\plotpoint}}
\put(533.13,493.81){\usebox{\plotpoint}}
\put(552.87,487.38){\usebox{\plotpoint}}
\put(572.30,480.13){\usebox{\plotpoint}}
\put(592.01,473.66){\usebox{\plotpoint}}
\put(611.70,467.10){\usebox{\plotpoint}}
\put(631.31,460.37){\usebox{\plotpoint}}
\put(650.86,453.43){\usebox{\plotpoint}}
\put(670.63,447.12){\usebox{\plotpoint}}
\put(690.32,440.56){\usebox{\plotpoint}}
\put(709.82,433.45){\usebox{\plotpoint}}
\put(729.49,426.84){\usebox{\plotpoint}}
\put(749.28,420.57){\usebox{\plotpoint}}
\put(768.73,413.36){\usebox{\plotpoint}}
\put(788.36,406.66){\usebox{\plotpoint}}
\put(808.11,400.30){\usebox{\plotpoint}}
\put(827.69,393.43){\usebox{\plotpoint}}
\put(847.28,386.57){\usebox{\plotpoint}}
\put(867.04,380.22){\usebox{\plotpoint}}
\put(886.66,373.48){\usebox{\plotpoint}}
\put(906.11,366.30){\usebox{\plotpoint}}
\put(925.90,360.03){\usebox{\plotpoint}}
\put(945.60,353.51){\usebox{\plotpoint}}
\put(965.15,346.60){\usebox{\plotpoint}}
\put(984.74,339.75){\usebox{\plotpoint}}
\put(1004.51,333.46){\usebox{\plotpoint}}
\put(1024.05,326.48){\usebox{\plotpoint}}
\put(1043.67,319.78){\usebox{\plotpoint}}
\put(1063.36,313.21){\usebox{\plotpoint}}
\put(1083.06,306.72){\usebox{\plotpoint}}
\put(1102.50,299.50){\usebox{\plotpoint}}
\put(1122.24,293.08){\usebox{\plotpoint}}
\put(1141.95,286.60){\usebox{\plotpoint}}
\put(1161.40,279.42){\usebox{\plotpoint}}
\put(1181.12,272.96){\usebox{\plotpoint}}
\put(1200.81,266.40){\usebox{\plotpoint}}
\put(1220.45,259.73){\usebox{\plotpoint}}
\put(1239.97,252.70){\usebox{\plotpoint}}
\put(1259.75,246.42){\usebox{\plotpoint}}
\put(1279.34,239.61){\usebox{\plotpoint}}
\put(1298.87,232.65){\usebox{\plotpoint}}
\put(1318.58,226.14){\usebox{\plotpoint}}
\put(1338.36,219.85){\usebox{\plotpoint}}
\put(1357.73,212.42){\usebox{\plotpoint}}
\put(1377.51,206.16){\usebox{\plotpoint}}
\put(1397.20,199.60){\usebox{\plotpoint}}
\put(1416.60,192.28){\usebox{\plotpoint}}
\put(1436,186){\usebox{\plotpoint}}
\end{picture}

%% file: fig8.tex
% GNUPLOT: LaTeX picture
\setlength{\unitlength}{0.240900pt}
\ifx\plotpoint\undefined\newsavebox{\plotpoint}\fi
\begin{picture}(1500,900)(0,0)
\font\gnuplot=cmr12 at 12pt
\gnuplot
\sbox{\plotpoint}{\rule[-0.200pt]{0.400pt}{0.400pt}}%
\put(241.0,134.0){\rule[-0.200pt]{4.818pt}{0.400pt}}
\put(219,134){\makebox(0,0)[r]{0.14}}
\put(1416.0,134.0){\rule[-0.200pt]{4.818pt}{0.400pt}}
\put(241.0,303.0){\rule[-0.200pt]{4.818pt}{0.400pt}}
\put(219,303){\makebox(0,0)[r]{0.145}}
\put(1416.0,303.0){\rule[-0.200pt]{4.818pt}{0.400pt}}
\put(241.0,472.0){\rule[-0.200pt]{4.818pt}{0.400pt}}
\put(219,472){\makebox(0,0)[r]{0.15}}
\put(1416.0,472.0){\rule[-0.200pt]{4.818pt}{0.400pt}}
\put(241.0,641.0){\rule[-0.200pt]{4.818pt}{0.400pt}}
\put(219,641){\makebox(0,0)[r]{0.155}}
\put(1416.0,641.0){\rule[-0.200pt]{4.818pt}{0.400pt}}
\put(241.0,810.0){\rule[-0.200pt]{4.818pt}{0.400pt}}
\put(219,810){\makebox(0,0)[r]{0.16}}
\put(1416.0,810.0){\rule[-0.200pt]{4.818pt}{0.400pt}}
\put(241.0,134.0){\rule[-0.200pt]{0.400pt}{4.818pt}}
\put(241,89){\makebox(0,0){0.001}}
\put(241.0,790.0){\rule[-0.200pt]{0.400pt}{4.818pt}}
\put(374.0,134.0){\rule[-0.200pt]{0.400pt}{4.818pt}}
\put(374,89){\makebox(0,0){0.002}}
\put(374.0,790.0){\rule[-0.200pt]{0.400pt}{4.818pt}}
\put(507.0,134.0){\rule[-0.200pt]{0.400pt}{4.818pt}}
\put(507,89){\makebox(0,0){0.003}}
\put(507.0,790.0){\rule[-0.200pt]{0.400pt}{4.818pt}}
\put(639.0,134.0){\rule[-0.200pt]{0.400pt}{4.818pt}}
\put(639,89){\makebox(0,0){0.004}}
\put(639.0,790.0){\rule[-0.200pt]{0.400pt}{4.818pt}}
\put(772.0,134.0){\rule[-0.200pt]{0.400pt}{4.818pt}}
\put(772,89){\makebox(0,0){0.005}}
\put(772.0,790.0){\rule[-0.200pt]{0.400pt}{4.818pt}}
\put(905.0,134.0){\rule[-0.200pt]{0.400pt}{4.818pt}}
\put(905,89){\makebox(0,0){0.006}}
\put(905.0,790.0){\rule[-0.200pt]{0.400pt}{4.818pt}}
\put(1038.0,134.0){\rule[-0.200pt]{0.400pt}{4.818pt}}
\put(1038,89){\makebox(0,0){0.007}}
\put(1038.0,790.0){\rule[-0.200pt]{0.400pt}{4.818pt}}
\put(1170.0,134.0){\rule[-0.200pt]{0.400pt}{4.818pt}}
\put(1170,89){\makebox(0,0){0.008}}
\put(1170.0,790.0){\rule[-0.200pt]{0.400pt}{4.818pt}}
\put(1303.0,134.0){\rule[-0.200pt]{0.400pt}{4.818pt}}
\put(1303,89){\makebox(0,0){0.009}}
\put(1303.0,790.0){\rule[-0.200pt]{0.400pt}{4.818pt}}
\put(1436.0,134.0){\rule[-0.200pt]{0.400pt}{4.818pt}}
\put(1436,89){\makebox(0,0){0.01}}
\put(1436.0,790.0){\rule[-0.200pt]{0.400pt}{4.818pt}}
\put(241.0,134.0){\rule[-0.200pt]{287.875pt}{0.400pt}}
\put(1436.0,134.0){\rule[-0.200pt]{0.400pt}{162.848pt}}
\put(241.0,810.0){\rule[-0.200pt]{287.875pt}{0.400pt}}
\put(45,472){\makebox(0,0){$u_1$}}
\put(45,428){\makebox(0,0){(arb. units)}}
\put(838,44){\makebox(0,0){$a_0$ (arb. units)}}
%\put(838,855){\makebox(0,0){Powerlaw exponent vs. lattice constant}}
\put(241.0,134.0){\rule[-0.200pt]{0.400pt}{162.848pt}}
\put(1262,768){\makebox(0,0)[r]{Data}}
\put(1338,768){\raisebox{-.8pt}{\makebox(0,0){$\diamond$}}}
\put(1436,736){\raisebox{-.8pt}{\makebox(0,0){$\diamond$}}}
\put(772,469){\raisebox{-.8pt}{\makebox(0,0){$\diamond$}}}
\put(550,401){\raisebox{-.8pt}{\makebox(0,0){$\diamond$}}}
\put(440,327){\raisebox{-.8pt}{\makebox(0,0){$\diamond$}}}
\put(374,325){\raisebox{-.8pt}{\makebox(0,0){$\diamond$}}}
\put(330,260){\raisebox{-.8pt}{\makebox(0,0){$\diamond$}}}
\put(298,328){\raisebox{-.8pt}{\makebox(0,0){$\diamond$}}}
\put(274,262){\raisebox{-.8pt}{\makebox(0,0){$\diamond$}}}
\put(256,381){\raisebox{-.8pt}{\makebox(0,0){$\diamond$}}}
\put(241,363){\raisebox{-.8pt}{\makebox(0,0){$\diamond$}}}
\put(1284.0,768.0){\rule[-0.200pt]{26.017pt}{0.400pt}}
\put(1284.0,758.0){\rule[-0.200pt]{0.400pt}{4.818pt}}
\put(1392.0,758.0){\rule[-0.200pt]{0.400pt}{4.818pt}}
\put(1436.0,694.0){\rule[-0.200pt]{0.400pt}{20.236pt}}
\put(1426.0,694.0){\rule[-0.200pt]{4.818pt}{0.400pt}}
\put(1426.0,778.0){\rule[-0.200pt]{4.818pt}{0.400pt}}
\put(772.0,451.0){\rule[-0.200pt]{0.400pt}{8.672pt}}
\put(762.0,451.0){\rule[-0.200pt]{4.818pt}{0.400pt}}
\put(762.0,487.0){\rule[-0.200pt]{4.818pt}{0.400pt}}
\put(550.0,378.0){\rule[-0.200pt]{0.400pt}{10.840pt}}
\put(540.0,378.0){\rule[-0.200pt]{4.818pt}{0.400pt}}
\put(540.0,423.0){\rule[-0.200pt]{4.818pt}{0.400pt}}
\put(440.0,315.0){\rule[-0.200pt]{0.400pt}{5.782pt}}
\put(430.0,315.0){\rule[-0.200pt]{4.818pt}{0.400pt}}
\put(430.0,339.0){\rule[-0.200pt]{4.818pt}{0.400pt}}
\put(374.0,305.0){\rule[-0.200pt]{0.400pt}{9.395pt}}
\put(364.0,305.0){\rule[-0.200pt]{4.818pt}{0.400pt}}
\put(364.0,344.0){\rule[-0.200pt]{4.818pt}{0.400pt}}
\put(330.0,237.0){\rule[-0.200pt]{0.400pt}{11.322pt}}
\put(320.0,237.0){\rule[-0.200pt]{4.818pt}{0.400pt}}
\put(320.0,284.0){\rule[-0.200pt]{4.818pt}{0.400pt}}
\put(298.0,310.0){\rule[-0.200pt]{0.400pt}{8.672pt}}
\put(288.0,310.0){\rule[-0.200pt]{4.818pt}{0.400pt}}
\put(288.0,346.0){\rule[-0.200pt]{4.818pt}{0.400pt}}
\put(274.0,245.0){\rule[-0.200pt]{0.400pt}{8.191pt}}
\put(264.0,245.0){\rule[-0.200pt]{4.818pt}{0.400pt}}
\put(264.0,279.0){\rule[-0.200pt]{4.818pt}{0.400pt}}
\put(256.0,360.0){\rule[-0.200pt]{0.400pt}{10.118pt}}
\put(246.0,360.0){\rule[-0.200pt]{4.818pt}{0.400pt}}
\put(246.0,402.0){\rule[-0.200pt]{4.818pt}{0.400pt}}
\put(241.0,349.0){\rule[-0.200pt]{0.400pt}{6.745pt}}
\put(231.0,349.0){\rule[-0.200pt]{4.818pt}{0.400pt}}
\put(231.0,377.0){\rule[-0.200pt]{4.818pt}{0.400pt}}
\put(1262,723){\makebox(0,0)[r]{Linear fit}}
\multiput(1284,723)(20.756,0.000){6}{\usebox{\plotpoint}}
\put(1392,723){\usebox{\plotpoint}}
\put(241,295){\usebox{\plotpoint}}
\put(241.00,295.00){\usebox{\plotpoint}}
\put(260.69,301.56){\usebox{\plotpoint}}
\put(280.29,308.37){\usebox{\plotpoint}}
\put(299.74,315.58){\usebox{\plotpoint}}
\put(319.43,322.14){\usebox{\plotpoint}}
\put(338.91,329.30){\usebox{\plotpoint}}
\put(358.60,335.87){\usebox{\plotpoint}}
\put(377.95,343.32){\usebox{\plotpoint}}
\put(397.64,349.88){\usebox{\plotpoint}}
\put(417.14,356.97){\usebox{\plotpoint}}
\put(436.69,363.90){\usebox{\plotpoint}}
\put(456.38,370.46){\usebox{\plotpoint}}
\put(475.74,377.91){\usebox{\plotpoint}}
\put(495.44,384.44){\usebox{\plotpoint}}
\put(515.00,391.33){\usebox{\plotpoint}}
\put(534.57,398.19){\usebox{\plotpoint}}
\put(554.26,404.75){\usebox{\plotpoint}}
\put(573.62,412.21){\usebox{\plotpoint}}
\put(593.25,418.94){\usebox{\plotpoint}}
\put(612.67,426.22){\usebox{\plotpoint}}
\put(632.36,432.79){\usebox{\plotpoint}}
\put(651.72,440.24){\usebox{\plotpoint}}
\put(671.47,446.61){\usebox{\plotpoint}}
\put(691.11,453.30){\usebox{\plotpoint}}
\put(710.55,460.52){\usebox{\plotpoint}}
\put(730.24,467.08){\usebox{\plotpoint}}
\put(749.60,474.53){\usebox{\plotpoint}}
\put(769.29,481.10){\usebox{\plotpoint}}
\put(788.85,488.02){\usebox{\plotpoint}}
\put(808.34,495.11){\usebox{\plotpoint}}
\put(827.81,502.26){\usebox{\plotpoint}}
\put(847.48,508.83){\usebox{\plotpoint}}
\put(867.17,515.39){\usebox{\plotpoint}}
\put(886.53,522.84){\usebox{\plotpoint}}
\put(906.22,529.41){\usebox{\plotpoint}}
\put(925.67,536.61){\usebox{\plotpoint}}
\put(945.27,543.42){\usebox{\plotpoint}}
\put(964.96,549.99){\usebox{\plotpoint}}
\put(984.32,557.44){\usebox{\plotpoint}}
\put(1004.03,563.93){\usebox{\plotpoint}}
\put(1023.53,570.97){\usebox{\plotpoint}}
\put(1043.15,577.72){\usebox{\plotpoint}}
\put(1062.82,584.34){\usebox{\plotpoint}}
\put(1082.20,591.73){\usebox{\plotpoint}}
\put(1101.79,598.58){\usebox{\plotpoint}}
\put(1121.25,605.75){\usebox{\plotpoint}}
\put(1140.94,612.31){\usebox{\plotpoint}}
\put(1160.30,619.77){\usebox{\plotpoint}}
\put(1180.06,626.10){\usebox{\plotpoint}}
\put(1199.64,632.94){\usebox{\plotpoint}}
\put(1219.13,640.04){\usebox{\plotpoint}}
\put(1238.82,646.61){\usebox{\plotpoint}}
\put(1258.18,654.06){\usebox{\plotpoint}}
\put(1277.87,660.62){\usebox{\plotpoint}}
\put(1297.38,667.66){\usebox{\plotpoint}}
\put(1316.92,674.64){\usebox{\plotpoint}}
\put(1336.35,681.90){\usebox{\plotpoint}}
\put(1356.06,688.35){\usebox{\plotpoint}}
\put(1375.75,694.92){\usebox{\plotpoint}}
\put(1395.11,702.37){\usebox{\plotpoint}}
\put(1414.80,708.93){\usebox{\plotpoint}}
\put(1434.21,716.25){\usebox{\plotpoint}}
\put(1436,717){\usebox{\plotpoint}}
\end{picture}

%% file: fig9.tex
% GNUPLOT: LaTeX picture
\setlength{\unitlength}{0.240900pt}
\ifx\plotpoint\undefined\newsavebox{\plotpoint}\fi
\sbox{\plotpoint}{\rule[-0.200pt]{0.400pt}{0.400pt}}%
\begin{picture}(1500,900)(0,0)
\font\gnuplot=cmr12 at 12pt
\gnuplot
\sbox{\plotpoint}{\rule[-0.200pt]{0.400pt}{0.400pt}}%
\put(219.0,134.0){\rule[-0.200pt]{4.818pt}{0.400pt}}
\put(197,134){\makebox(0,0)[r]{0.1}}
\put(1416.0,134.0){\rule[-0.200pt]{4.818pt}{0.400pt}}
\put(219.0,269.0){\rule[-0.200pt]{4.818pt}{0.400pt}}
\put(197,269){\makebox(0,0)[r]{0.12}}
\put(1416.0,269.0){\rule[-0.200pt]{4.818pt}{0.400pt}}
\put(219.0,404.0){\rule[-0.200pt]{4.818pt}{0.400pt}}
\put(197,404){\makebox(0,0)[r]{0.14}}
\put(1416.0,404.0){\rule[-0.200pt]{4.818pt}{0.400pt}}
\put(219.0,540.0){\rule[-0.200pt]{4.818pt}{0.400pt}}
\put(197,540){\makebox(0,0)[r]{0.16}}
\put(1416.0,540.0){\rule[-0.200pt]{4.818pt}{0.400pt}}
\put(219.0,675.0){\rule[-0.200pt]{4.818pt}{0.400pt}}
\put(197,675){\makebox(0,0)[r]{0.18}}
\put(1416.0,675.0){\rule[-0.200pt]{4.818pt}{0.400pt}}
\put(219.0,810.0){\rule[-0.200pt]{4.818pt}{0.400pt}}
\put(197,810){\makebox(0,0)[r]{0.2}}
\put(1416.0,810.0){\rule[-0.200pt]{4.818pt}{0.400pt}}
\put(219.0,134.0){\rule[-0.200pt]{0.400pt}{4.818pt}}
\put(219,89){\makebox(0,0){10}}
\put(219.0,790.0){\rule[-0.200pt]{0.400pt}{4.818pt}}
\put(354.0,134.0){\rule[-0.200pt]{0.400pt}{4.818pt}}
\put(354,89){\makebox(0,0){20}}
\put(354.0,790.0){\rule[-0.200pt]{0.400pt}{4.818pt}}
\put(489.0,134.0){\rule[-0.200pt]{0.400pt}{4.818pt}}
\put(489,89){\makebox(0,0){30}}
\put(489.0,790.0){\rule[-0.200pt]{0.400pt}{4.818pt}}
\put(625.0,134.0){\rule[-0.200pt]{0.400pt}{4.818pt}}
\put(625,89){\makebox(0,0){40}}
\put(625.0,790.0){\rule[-0.200pt]{0.400pt}{4.818pt}}
\put(760.0,134.0){\rule[-0.200pt]{0.400pt}{4.818pt}}
\put(760,89){\makebox(0,0){50}}
\put(760.0,790.0){\rule[-0.200pt]{0.400pt}{4.818pt}}
\put(895.0,134.0){\rule[-0.200pt]{0.400pt}{4.818pt}}
\put(895,89){\makebox(0,0){60}}
\put(895.0,790.0){\rule[-0.200pt]{0.400pt}{4.818pt}}
\put(1030.0,134.0){\rule[-0.200pt]{0.400pt}{4.818pt}}
\put(1030,89){\makebox(0,0){70}}
\put(1030.0,790.0){\rule[-0.200pt]{0.400pt}{4.818pt}}
\put(1166.0,134.0){\rule[-0.200pt]{0.400pt}{4.818pt}}
\put(1166,89){\makebox(0,0){80}}
\put(1166.0,790.0){\rule[-0.200pt]{0.400pt}{4.818pt}}
\put(1301.0,134.0){\rule[-0.200pt]{0.400pt}{4.818pt}}
\put(1301,89){\makebox(0,0){90}}
\put(1301.0,790.0){\rule[-0.200pt]{0.400pt}{4.818pt}}
\put(1436.0,134.0){\rule[-0.200pt]{0.400pt}{4.818pt}}
\put(1436,89){\makebox(0,0){100}}
\put(1436.0,790.0){\rule[-0.200pt]{0.400pt}{4.818pt}}
\put(219.0,134.0){\rule[-0.200pt]{293.175pt}{0.400pt}}
\put(1436.0,134.0){\rule[-0.200pt]{0.400pt}{162.848pt}}
\put(219.0,810.0){\rule[-0.200pt]{293.175pt}{0.400pt}}
\put(45,492){\makebox(0,0){$\Delta x$}}
\put(45,448){\makebox(0,0){(arb. units)}}
\put(827,44){\makebox(0,0){$v_0$ (arb. units)}}
%\put(827,855){\makebox(0,0){Width of distribution as a function of initial speed}}
\put(219.0,134.0){\rule[-0.200pt]{0.400pt}{162.848pt}}
\put(1262,768){\makebox(0,0)[r]{Data}}
\put(1338,768){\raisebox{-.8pt}{\makebox(0,0){$\diamond$}}}
\put(219,458){\raisebox{-.8pt}{\makebox(0,0){$\diamond$}}}
\put(354,472){\raisebox{-.8pt}{\makebox(0,0){$\diamond$}}}
\put(489,452){\raisebox{-.8pt}{\makebox(0,0){$\diamond$}}}
\put(625,484){\raisebox{-.8pt}{\makebox(0,0){$\diamond$}}}
\put(760,496){\raisebox{-.8pt}{\makebox(0,0){$\diamond$}}}
\put(895,461){\raisebox{-.8pt}{\makebox(0,0){$\diamond$}}}
\put(1030,480){\raisebox{-.8pt}{\makebox(0,0){$\diamond$}}}
\put(1166,473){\raisebox{-.8pt}{\makebox(0,0){$\diamond$}}}
\put(1301,483){\raisebox{-.8pt}{\makebox(0,0){$\diamond$}}}
\put(1436,488){\raisebox{-.8pt}{\makebox(0,0){$\diamond$}}}
\put(1262,723){\makebox(0,0)[r]{Average value}}
\multiput(1284,723)(20.756,0.000){6}{\usebox{\plotpoint}}
\put(1392,723){\usebox{\plotpoint}}
\put(219,475){\usebox{\plotpoint}}
\put(219.00,475.00){\usebox{\plotpoint}}
\put(239.76,475.00){\usebox{\plotpoint}}
\put(260.51,475.00){\usebox{\plotpoint}}
\put(281.27,475.00){\usebox{\plotpoint}}
\put(302.02,475.00){\usebox{\plotpoint}}
\put(322.78,475.00){\usebox{\plotpoint}}
\put(343.53,475.00){\usebox{\plotpoint}}
\put(364.29,475.00){\usebox{\plotpoint}}
\put(385.04,475.00){\usebox{\plotpoint}}
\put(405.80,475.00){\usebox{\plotpoint}}
\put(426.55,475.00){\usebox{\plotpoint}}
\put(447.31,475.00){\usebox{\plotpoint}}
\put(468.07,475.00){\usebox{\plotpoint}}
\put(488.82,475.00){\usebox{\plotpoint}}
\put(509.58,475.00){\usebox{\plotpoint}}
\put(530.33,475.00){\usebox{\plotpoint}}
\put(551.09,475.00){\usebox{\plotpoint}}
\put(571.84,475.00){\usebox{\plotpoint}}
\put(592.60,475.00){\usebox{\plotpoint}}
\put(613.35,475.00){\usebox{\plotpoint}}
\put(634.11,475.00){\usebox{\plotpoint}}
\put(654.87,475.00){\usebox{\plotpoint}}
\put(675.62,475.00){\usebox{\plotpoint}}
\put(696.38,475.00){\usebox{\plotpoint}}
\put(717.13,475.00){\usebox{\plotpoint}}
\put(737.89,475.00){\usebox{\plotpoint}}
\put(758.64,475.00){\usebox{\plotpoint}}
\put(779.40,475.00){\usebox{\plotpoint}}
\put(800.15,475.00){\usebox{\plotpoint}}
\put(820.91,475.00){\usebox{\plotpoint}}
\put(841.66,475.00){\usebox{\plotpoint}}
\put(862.42,475.00){\usebox{\plotpoint}}
\put(883.18,475.00){\usebox{\plotpoint}}
\put(903.93,475.00){\usebox{\plotpoint}}
\put(924.69,475.00){\usebox{\plotpoint}}
\put(945.44,475.00){\usebox{\plotpoint}}
\put(966.20,475.00){\usebox{\plotpoint}}
\put(986.95,475.00){\usebox{\plotpoint}}
\put(1007.71,475.00){\usebox{\plotpoint}}
\put(1028.46,475.00){\usebox{\plotpoint}}
\put(1049.22,475.00){\usebox{\plotpoint}}
\put(1069.98,475.00){\usebox{\plotpoint}}
\put(1090.73,475.00){\usebox{\plotpoint}}
\put(1111.49,475.00){\usebox{\plotpoint}}
\put(1132.24,475.00){\usebox{\plotpoint}}
\put(1153.00,475.00){\usebox{\plotpoint}}
\put(1173.75,475.00){\usebox{\plotpoint}}
\put(1194.51,475.00){\usebox{\plotpoint}}
\put(1215.26,475.00){\usebox{\plotpoint}}
\put(1236.02,475.00){\usebox{\plotpoint}}
\put(1256.77,475.00){\usebox{\plotpoint}}
\put(1277.53,475.00){\usebox{\plotpoint}}
\put(1298.29,475.00){\usebox{\plotpoint}}
\put(1319.04,475.00){\usebox{\plotpoint}}
\put(1339.80,475.00){\usebox{\plotpoint}}
\put(1360.55,475.00){\usebox{\plotpoint}}
\put(1381.31,475.00){\usebox{\plotpoint}}
\put(1402.06,475.00){\usebox{\plotpoint}}
\put(1422.82,475.00){\usebox{\plotpoint}}
\put(1436,475){\usebox{\plotpoint}}
\end{picture}